\documentclass[fleqn,usenatbib]{mnras}

\usepackage{newtxtext,newtxmath}

\usepackage[T1]{fontenc}
\usepackage{ae,aecompl}

\usepackage{graphicx}	
\usepackage{amsmath}	
\usepackage{amssymb}	

\usepackage{soul}
\usepackage[dvipsnames]{xcolor}

\usepackage{etoolbox}
\makeatletter
\patchcmd\@combinedblfloats{\box\@outputbox}{\unvbox\@outputbox}{}{%
  \errmessage{\noexpand\@combinedblfloats could not be patched}%
}%
\makeatother

\newcommand{\Msun}{\textrm{M}_\odot}

\defcitealias{Hafen2019a}{H19}
\newcommand{\originp}{\citetalias{Hafen2019a}}

\title[Fates of the CGM]{The Fates of the Circumgalactic Medium in the FIRE Simulations}

\author[Hafen et al.]{
\parbox{\textwidth}{
Zachary Hafen,$^{1}$\thanks{E-mail: zhafen@u.northwestern.edu}
Claude-Andr\'e Faucher-Gigu\`ere,$^{1}$
Daniel Angl\'es-Alc\'azar,$^{2,3}$
Jonathan~Stern,$^1$
Du\v{s}an Kere\v{s},$^4$
Clarke Esmerian,$^5$
Andrew Wetzel,$^6$
Kareem El-Badry,$^7$
T. K. Chan,$^{4,8}$
Norman Murray$^{9\thanks{Canada Research Chair in Astrophysics.}}$
}
\vspace{0.4cm} \\
$^{1}$Department of Physics and Astronomy and Center for Interdisciplinary Exploration and Research in Astrophysics (CIERA),\\ Northwestern University, 2145 Sheridan Road, Evanston, IL 60208, USA \\
$^{2}$Center for Computational Astrophysics, Flatiron Institute, 162 Fifth Avenue, New York, NY 10010, USA\\
$^3$Department of Physics, University of Connecticut, 196 Auditorium Road, U-3046, Storrs, CT 06269-3046, USA\\
$^4$Department of Physics, Center for Astrophysics and Space Sciences, University of California, San Diego, 9500 Gilman Drive, \\ La Jolla, CA 9209, USA \\
$^5$Department of Astronomy and Astrophysics, The University of Chicago, Chicago, IL 60637, USA \\
$^6$Department of Physics, University of California, Davis, CA, USA 95616 \\
$^7$Department of Astronomy and Theoretical Astrophysics Center, University of California, Berkeley, CA 94720-3411, USA \\
$^8$Institute for Computational Cosmology, Durham University, South Road, Durham, DH1 3LE, UK \\
$^9$Canadian Institute for Theoretical Astrophysics, 60 St George Street, University of Toronto, ON M5S 3H8, Canada
}

\date{Accepted XXX. Received YYY; in original form ZZZ}

\pubyear{2015}

\begin{document}
\label{firstpage}
\pagerange{\pageref{firstpage}--\pageref{lastpage}}
\maketitle

\newcommand{\percentejectedMWmasslow}{5\%}
\newcommand{\percentaccretedminmasslow}{15\%}
\newcommand{\percentejectedMWmasshigh}{30\%}
\newcommand{\percentaccretedminmasshigh}{25\%}
\newcommand{\accdivgalmedianmasshigh}{1.6}
\newcommand{\accdivgalmedianmasslow}{0.5}
\newcommand{\accsatmaxmasslow}{25\%}
\newcommand{\stillCGMmedmasslow}{30\%}
\newcommand{\percentejectedMWmetalmasslow}{5\%}
\newcommand{\percentaccretedminmetalmasslow}{20\%}
\newcommand{\accdivgalmedianmetalmasslow}{0.54}
\newcommand{\accsatmaxmetalmasslow}{40\%}
\newcommand{\stillCGMmedmetalmasslow}{20\%}
\newcommand{\percentejectedMWmetalmasshigh}{35\%}
\newcommand{\percentaccretedminmetalmasshigh}{35\%}
\newcommand{\accdivgalmedianmetalmasshigh}{1.4}
\newcommand{\accsatmaxmetalmasshigh}{25\%}
\newcommand{\stillCGMmedmetalmasshigh}{0\%}
\newcommand{\accsatmaxmasshigh}{25\%}
\newcommand{\stillCGMmedmasshigh}{0\%}
\newcommand{\percentaccretedIGMaccminustwosigmamasslow}{35}
\newcommand{\percentaccretedIGMaccplustwosigmamasslow}{60\%}
\newcommand{\percentIGMaccaccretedminustwosigmamasslow}{15}
\newcommand{\percentIGMaccaccretedplustwosigmamasslow}{40\%}
\newcommand{\percentwindaccretedminustwosigmamasslow}{25}
\newcommand{\percentwindaccretedplustwosigmamasslow}{60\%}
\newcommand{\percentsatwindaccretedminustwosigmamasslow}{15}
\newcommand{\percentsatwindaccretedplustwosigmamasslow}{45\%}
\newcommand{\percentIGMaccejectedminustwosigmamasslow}{5}
\newcommand{\percentIGMaccejectedplustwosigmamasslow}{40\%}
\newcommand{\percentIGMaccejectedminustwosigmamasshigh}{20}
\newcommand{\percentIGMaccejectedplustwosigmamasshigh}{60\%}
\newcommand{\percentIGMaccaccretedminustwosigmamasshigh}{25}
\newcommand{\percentIGMaccaccretedplustwosigmamasshigh}{75\%}
\newcommand{\percentwindaccretedminustwosigmamasshigh}{40}
\newcommand{\percentwindaccretedplustwosigmamasshigh}{65\%}
\newcommand{\percentsatwindaccretedminustwosigmamasshigh}{30}
\newcommand{\percentsatwindaccretedplustwosigmamasshigh}{80\%}
\newcommand{\percentaccretedIGMaccminustwosigmamasshigh}{45}
\newcommand{\percentaccretedIGMaccplustwosigmamasshigh}{55\%}
\newcommand{\percentaccsatsatwindminustwosigmamasshigh}{5}
\newcommand{\percentaccsatsatwindplustwosigmamasshigh}{50\%}
\newcommand{\percentaccsatsatwindminustwosigmamasslow}{60}
\newcommand{\percentaccsatsatwindplustwosigmamasslow}{80\%}
\newcommand{\percentaccsatandsatwindmasslow}{5\%}
\newcommand{\percentaccsatandsatwindmasshigh}{0\%}
\newcommand{\percentaccsatsatwindmedmasshigh}{15\%}
\newcommand{\percentaccsatsatwindmedmasslow}{75\%}
\newcommand{\Fordcomphighmassmasslow}{40\%}
\newcommand{\Fordcomplowmassmasslow}{30\%}
\newcommand{\lowratioofmedmetacctomedmetstilllow}{4}
\newcommand{\medratioofmedmetacctomedmetstilllow}{10}
\newcommand{\inCGMtoCGMstillmedmasslow}{7\%}
\newcommand{\inCGMtoCGMstillmedmasshigh}{11\%}
\newcommand{\Fordcomphighmassmasshigh}{60\%}
\newcommand{\Fordcomplowmassmasshigh}{60\%}
\newcommand{\percentIGMaccejectedminustwosigmametalmasshigh}{25}
\newcommand{\percentIGMaccejectedplustwosigmametalmasshigh}{60\%}
\newcommand{\percentIGMaccaccretedminustwosigmametalmasshigh}{30}
\newcommand{\percentIGMaccaccretedplustwosigmametalmasshigh}{70\%}
\newcommand{\percentwindaccretedminustwosigmametalmasshigh}{40}
\newcommand{\percentwindaccretedplustwosigmametalmasshigh}{65\%}
\newcommand{\percentsatwindaccretedminustwosigmametalmasshigh}{30}
\newcommand{\percentsatwindaccretedplustwosigmametalmasshigh}{80\%}
\newcommand{\percentaccretedIGMaccminustwosigmametalmasshigh}{0}
\newcommand{\percentaccretedIGMaccplustwosigmametalmasshigh}{15\%}
\newcommand{\percentaccsatsatwindminustwosigmametalmasshigh}{5}
\newcommand{\percentaccsatsatwindplustwosigmametalmasshigh}{50\%}
\newcommand{\percentaccsatsatwindmedmetalmasshigh}{20\%}
\newcommand{\inCGMtoCGMstillmedmetalmasshigh}{102\%}
\newcommand{\Fordcomphighmassmetalmasshigh}{60\%}
\newcommand{\Fordcomplowmassmetalmasshigh}{50\%}
\newcommand{\percentIGMaccejectedminustwosigmametalmasslow}{5}
\newcommand{\percentIGMaccejectedplustwosigmametalmasslow}{30\%}
\newcommand{\percentIGMaccaccretedminustwosigmametalmasslow}{25}
\newcommand{\percentIGMaccaccretedplustwosigmametalmasslow}{60\%}
\newcommand{\percentwindaccretedminustwosigmametalmasslow}{30}
\newcommand{\percentwindaccretedplustwosigmametalmasslow}{70\%}
\newcommand{\percentsatwindaccretedminustwosigmametalmasslow}{10}
\newcommand{\percentsatwindaccretedplustwosigmametalmasslow}{50\%}
\newcommand{\percentaccretedIGMaccminustwosigmametalmasslow}{0}
\newcommand{\percentaccretedIGMaccplustwosigmametalmasslow}{20\%}
\newcommand{\percentaccsatsatwindminustwosigmametalmasslow}{55}
\newcommand{\percentaccsatsatwindplustwosigmametalmasslow}{90\%}
\newcommand{\percentaccsatsatwindmedmetalmasslow}{80\%}
\newcommand{\inCGMtoCGMstillmedmetalmasslow}{34\%}
\newcommand{\Fordcomphighmassmetalmasslow}{60\%}
\newcommand{\Fordcomplowmassmetalmasslow}{50\%}

\begin{abstract}
We analyze the different fates of the circumgalactic medium (CGM) in FIRE-2 cosmological simulations, focusing on the redshifts $z=0.25$ and $z=2$ representative of recent surveys.
Our analysis includes 21 zoom-in simulations covering the halo mass range $M_{\rm h}(z=0) \sim 10^{10} - 10^{12} \Msun$. 
We analyze both where the gas ends up after first leaving the CGM (its ``proximate'' fate), as well as its location at $z=0$ (its ``ultimate'' fate). 
Of the CGM at $z=2$, about half is found in the ISM or stars of the central galaxy by $z=0$ in $M_{\rm h}(z=2) \sim 5\times10^{11}$ M$_{\odot}$ halos, but most of the CGM in lower-mass halos ends up in the IGM.
This is so even though most of the CGM in $M_{\rm h}(z=2) \sim 5\times10^{10}$ M$_{\odot}$ halos first accretes onto the central galaxy before being ejected into the IGM.
On the other hand, most of the CGM mass at $z=0.25$ remains in the CGM by $z=0$ at all halo masses analyzed.
Of the CGM gas that subsequently accretes onto the central galaxy in the progenitors of $M_{\rm h}(z=0)\sim10^{12}$ M$_{\odot}$ halos, most of it is cool ($T\sim10^{4}$ K) at $z=2$ but hot ($\sim T_{\rm vir}$) at $z\sim0.25$, consistent with the expected transition from cold mode to hot mode accretion.
Despite the transition in accretion mode, at both $z=0.25$ and $z=2$ $\gtrsim 80\%$ of the cool gas in $M_{\rm h} \gtrsim 10^{11} \Msun$ halos will accrete onto a galaxy. 
We find that the metallicity of CGM gas is typically a poor predictor of both its proximate and ultimate fates.
This is because there is in general little correlation between the origin of CGM gas and its fate owing to substantial mixing while in the CGM.
\end{abstract}




\section{Introduction}
\label{sec:intro}

Both observations and simulations indicate that galaxies from dwarfs through galaxies in clusters are enclosed in enormous gaseous atmospheres~\citep[e.g.][]{Steidel2010, Hummels2013, Ford2013, Bordoloi2014a, Werk2014, Liang2014, Stocke2014, Johnson2017}.
It has become clear in recent years that these gaseous atmospheres, also known as the circumgalactic medium (CGM), crucially affect the evolution of galaxies. 
For example, the CGM mass can exceed that of the central galaxy ~\citep[e.g][]{Tumlinson2017, Hafen2019a}. 
Moreover, the CGM mediates powerful interactions between galaxies and the larger-scale intergalactic medium (IGM), such as cosmic inflows~\citep[e.g.][]{Keres2005, Dekel2006, Prochaska2009, Faucher-Giguere2011a} and powerful galactic winds~\citep[e.g.][]{Martin2005, Veilleux2005, Dave2011, Muratov2015, Fielding2016, Angles-Alcazar2017}.

The large gas mass present in the CGM, much of which is enriched with metals, raises the question of what happens to the CGM observed in different surveys. 
I.e, what are the different fates of the CGM? 
This a key question for the physical interpretation of observational surveys of the CGM, at both low \citep[e.g.][]{Tumlinson2011, Stocke2013, Bouche2013, Bordoloi2014a, Rubin2014, Johnson2015, Burchett2018a, Chen2018} and high \citep[e.g.][]{Steidel2010, Rudie2012, Prochaska2013} redshift. 
The fates of the CGM, in particular, inform us about the whether circumgalactic gas primarily accretes onto galaxies, moves out in outflows that potentially enrich the IGM with metals, or remains in the CGM for cosmological timescales. 
The fates of the CGM are also important for our understanding of different observed phenomena. 
Consider the massive CGM of low-redshift $\sim L^{\star}$ galaxies, which have been the focus of comprehensive surveys with HST/COS \citep[e.g.,][]{Werk2013, Johnson2015, Keeney2017}. 
If this CGM remains as gaseous halos until $z \approx 0$, then it may contribute to observed high-velocity clouds~\citep[e.g.][]{Putman2012} or X-ray emitting gas~\citep[e.g.][]{Henley2010, Henley2010a, Gupta2012, Fang2015}. 
On the other hand, if all the \ion{O}{vi}-traced gas in $L^\star$ halos accretes onto the central galaxy, then it could provide fuel for stars to form at a rate up to $\sim 1-10\times$ the observed star-formation rate~\citep[][]{Mathews2017, McQuinn2018, Stern2018}. 

In this work, we trace the history of gas elements (commonly referred to as ``particle tracking'') to study the different fates of the CGM in the FIRE cosmological zoom-in simulations \citep[][]{Hopkins2014,Hopkins2017}.\footnote{FIRE project website: \url{https://fire.northwestern.edu/}}
This paper complements our previous analysis of the origins of the CGM in the same simulations~\citep[][hereafter H19]{Hafen2019a}, as well as our analysis of the role of the cosmic baryon cycle in the build up of central galaxies in FIRE \citep[][]{Angles-Alcazar2017}. 
The present analysis also complements several previous particle tracking studies in other simulations, which generally adopted a ``galaxy-centric'' rather than a ''CGM-centric'' perspective \citep[e.g.][]{Oppenheimer2010, VandeVoort2011, Nelson2013, Ford2014, Christensen2018, Tollet2019, Ho2019}. 
The FIRE simulations simultaneously include the cosmological environment as well as the effects of galactic winds generated by energy injection on the scale of individual star-forming regions, making them well suited for predictions of the gas flows that occur in the CGM.
In previous papers, we analyzed other bulk properties of the CGM in FIRE simulations~\citep{Muratov2015, Muratov2016, Stewart2016, El-Badry2017, Ji2019} and have made predictions of its observability in absorption and emission \citep[][]{Faucher-Giguere2015, Faucher-Giguere2016, Sravan2016, Hafen2016, VandeVoort2016}. 

Our analysis covers both the fate of gas immediately after it leaves the CGM (which we refer to as its \textit{proximate fate}) and the fate of the gas by $z=0$ (which we refer to as its \textit{ultimate fate}).
By studying the proximate fates of CGM gas, we address how CGM gas elements contribute to the baryon cycle in the near term.
Our analysis of the ultimate fates of the CGM tells us where the gas ends up by $z=0$. 
Analyzing both the proximate and ultimate fates is important to gain insight into the baryon cycle because of the importance of recycling and ``fall back'': gas which accretes onto a galaxy can later be ejected back into the CGM by galactic winds, while gas which ejected from halos can later reaccrete onto the CGM. 
We focus on the CGM at $z=0.25$ and $z=2$, which are representative of major recent surveys referenced above.

The structure of this paper is as follows. 
In \S\ref{sec:methods} we describe the simulations used and our particle tracking analysis methods.
Our main results are presented in \S\ref{sec:results}, including the past and future locations of CGM gas (\S\ref{sec:baryon_cycle}), trajectories of gas elements of different fates (\S\ref{sec:particle_pathlines}), and how CGM fates depend on halo mass, temperature, and metallicity (\S\ref{sec:fates_of_the_CGM}). 
We also address the connection between the origin of a CGM gas element and its fate (\S\ref{sec:origin_fate_connection}). 
We discuss our results in \S\ref{sec:discussion} and conclude in \S\ref{sec:conclusion}. 

Throughout, we assume a standard flat $\Lambda$CDM cosmology with $\Omega_{\rm m }\approx 0.32$, $\Omega_{\Lambda}=1-\Omega_{\rm m}$, $\Omega_{\rm b} \approx 0.049$, and $H_{0} \approx 67$ km s$^{-1}$ Mpc$^{-1}$ \citep[][]{PlanckCollaboration2018}.
\footnote{For consistency with previous work, some of our simulations were evolved with slightly different sets of cosmological parameters, but we do not expect this to significantly impact on any of our results given the small differences in the parameters.}

\section{Methods}
\label{sec:methods}

\subsection{Simulations}
\label{sec:simulations}

Our analysis is performed on a sample of FIRE-2 cosmological hydrodynamic ``zoom-in'' simulations previously analyzed in \originp, where they are described more fully.
The simulations were produced with the multi-method gravity and hydrodynamics code \textsc{GIZMO}\footnote{\url{http://www.tapir.caltech.edu/\~phopkins/Site/GIZMO.html}}~\citep{Hopkins2015} in its meshless finite-mass (``MFM'') mode.
In MFM there is no mass flux between resolution elements, which allows us to follow the history of gas by following resolution elements.
We analyze 21 simulations with main halos (listed in Table 1 of \originp) spanning the halo mass range of $M_{\rm h}(z=0) \sim 10^{10} - 10^{12} \Msun$. 
As in \originp, for the sake of brevity we refer to main halos with $M_{\rm h}(z=0) \sim 10^{10}, 10^{11}, 10^{12} \Msun$ simply as $10^{10} \Msun$ progenitors,  $10^{11} \Msun$ progenitors, and $10^{12} \Msun$ progenitors
(these also correspond to simulations whose name begins with \texttt{m10}, \texttt{m11}, or \texttt{m12}, respectively).  
We identify and track the evolution of dark matter halos using the Amiga Halo Finder~\citep[AHF;][]{Gill2004,Knollmann2009} and adopt the virial halo overdensity definition of \cite{Bryan1998}.

We summarize the key elements of our simulation method here; full details are provided in \cite{Hopkins2017}.
The simulations follow radiative heating and cooling over  $T=10 - 10^{10}$ K,  and include the effects of free-free emission, Compton scattering with the cosmic microwave background, photoelectric heating, high-temperature metal line cooling, and approximations for low-temperature cooling by molecules  and fine-structure metal lines.
Photo-heating and photo-ionization include effects of both a cosmic UV background model \citep{Faucher-Giguere2009} as well as approximations for local sources and self-shielding of dense gas.
Star formation occurs only in self-gravitating gas, identified with the criteria of \citealt{Hopkins2013b}.
In addition, we require that star formation only occurs in gas that it is molecular, self-shielding, and has a density of at least $n_{\rm H} = 1000$ cm$^{-3}$.
Stellar feedback includes momentum from radiation pressure; energy, momentum, mass, and metals from Type Ia and II supernovae and stellar winds; and photo-ionization and photo-electric heating.
Star particles are treated as independent stellar populations, with feedback quantities drawn from the \textsc{STARBURST99} stellar evolution models~\citep{Leitherer1999} assuming the IMF of \cite{Kroupa2001}.
We track the evolution of 9 independent metal species.

Of the 21 simulations analyzed in this work, 9 were produced with a sub-grid model for metal transport between adjacent resolution elements.
The purpose of the model is to capture sub-resolution-level metal diffusion that is not by default accounted for by the MFM hydrodynamic solver. 
The primary assumption for the metal diffusion model is that the diffusion timescale scales with the eddy turnover time of the largest unresolved turbulent eddies, i.e. that unresolved turbulence exchanges metals between resolution elements.
The details of our metal transport model are described in \cite{Hopkins2017a}, \cite{Hopkins2017}, and \cite{Escala2018}.

\subsection{Baryon Cycle Definitions}
\label{sec:baryon_cycle_defs}

During the baryon cycle, baryons can be found in the IGM, the CGM, or galaxies. 
Baryons in galaxies can be either in the interstellar medium (ISM) or in stars.
The definitions for galaxies and the CGM are the same as in \originp.
To summarize, for a given halo all gas within a radius $R_{\rm gal}$ and with $n_{\rm H} > 0.13$ cm$^{-3}$ is considered part of the central galaxy's ISM and all stars within $R_{\rm gal}$ are considered part of that galaxy. 
We use $R_{\rm gal} = 4R_{\star,0.5}$, where $R_{\star,0.5}$ is the stellar half-mass radius.
For each halo the CGM is defined as all the gas outside both the central galaxy and satellite galaxies, and in the case of the simulation's main halo we further require that the CGM gas has $r>R_{\rm CGM,~inner}$ where $R_{\rm CGM,~inner} \equiv \max( 1.2 R_{\rm gal}, 0.1 R_{\rm vir})$.
\footnote{
In \originp~we included satellite ISM as part of the CGM for the purpose of quantifying its mass contribution to the CGM.
Having determined that satellite ISM is $\lesssim 5\%$ of the CGM mass, we do not consider satellite ISM part of the CGM in this paper.
}
This definition allows us to focus our analysis on CGM gas that is clearly separate from the main galaxy. 
The remaining main halo gas, i.e. gas within $R_{\rm gal}$ but $n_{\rm H} < 0.13$ cm$^{-3}$ and gas with $R_{\rm gal} < r < R_{\rm CGM,inner}$, is defined as the ``galaxy-halo interface.'' 
We define the IGM as gas outside the virial radii of all halos in a simulation. 

Zoom-in simulations simulate a single main halo at high-resolution, and for each zoom-in we focus our analysis on the CGM of the main halo.
We thus differentiate halos that are satellites of the main halo and halos external to the main halo. 
We do not distinguish the CGM of satellites from that of the main halo, given the extent to which they can mix through stripping and other processes.
For the same reason we only identify the galaxy-halo interface of the main galaxy, and not of satellite galaxies.

\subsection{Proximate and Ultimate Fate Classifications}
\label{sec:classifications}
\begin{figure}
\centering
\includegraphics[width=0.75\columnwidth]{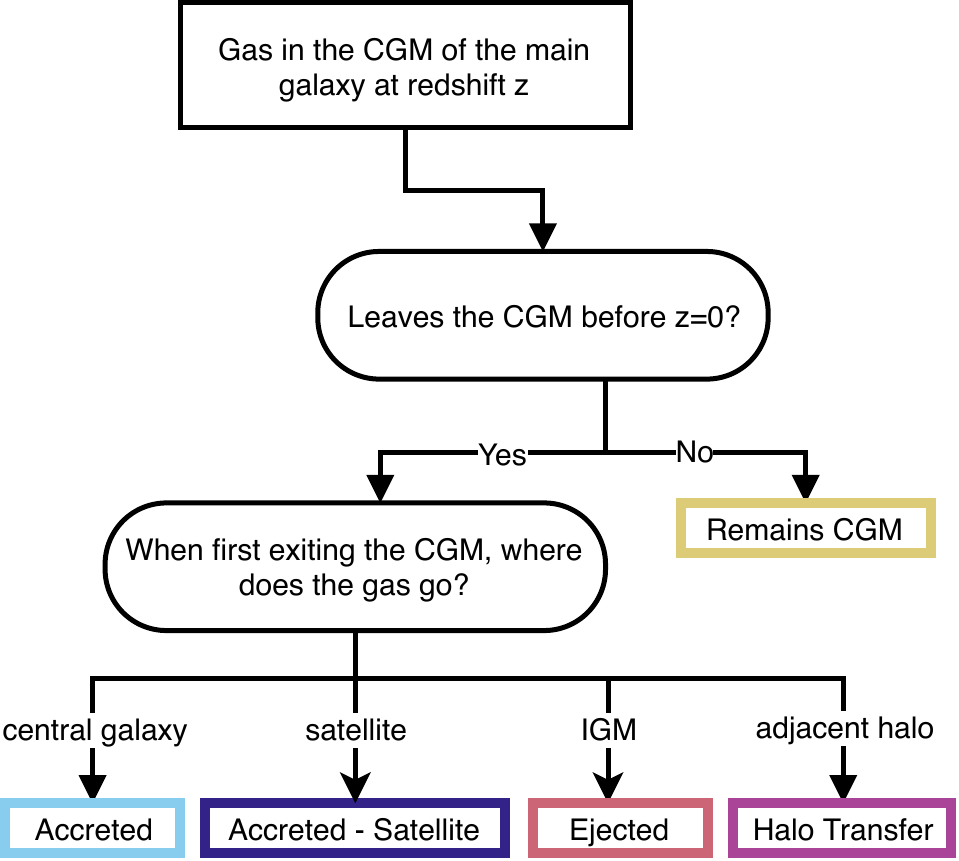}
\caption{
Flow chart summarizing how we classify the proximate fate of gas elements in the CGM of a galaxy.
Gas is classified as \textit{accreted} (onto the central galaxy), \textit{accreted-satellite} (accreted onto a satellite), \textit{ejected} (into the IGM), or \textit{halo transfer} depending on the component of the baryon cycle it transfers to upon leaving the CGM.
Some gas stays continuously in the CGM until $z=0$, i.e. it \textit{remains CGM}.
}
\label{fig:CGM_fates}
\end{figure}

The sample of particles tracked in this work is the same as in \originp.
Briefly, at both $z=0.25$ and $z=2$ we randomly sample $10^5$ particles in the CGM of each main halo.
We compile the full history of these particles at $\sim 25$ Myr time resolution, and use those histories to assign a proximate fate and an ultimate fate to each particle.
Note that the proximate fate in general depends on the detailed history of a gas element, while the ultimate fate only depends on the final location of the gas.

The ultimate fate classification for a particle is simply the component of the baryon cycle the particle resides in at $z=0$. 
This can be: the main galaxy's CGM, the main galaxy, the galaxy-halo interface, a satellite galaxy, an external galaxy, the CGM of an external galaxy, or the IGM.

The proximate fate of CGM gas identifies the next ``component transition'' experienced. 
The main proximate fates are: accretion onto the central galaxy, accretion onto a satellite galaxy, ejection into the IGM, or continuous residence in the CGM until $z=0$.
Another possible proximate fate is ``halo transfer.'' 
This occurs when gas is found within the halo of an external galaxy the snapshot after it was last in the CGM.
We have quantified halo transfer, but found that $\lesssim 1\%$ of the CGM mass contributes to it, so we omit it from the figures in this paper for simplicity.
Figure~\ref{fig:CGM_fates} uses a flow chart to summarize how we assign proximate fates.
Note that the proximate fates are insensitive to the timescale on which gas leaves the CGM, with the exception of gas that remains in the CGM until $z=0$.

Some care is needed to robustly identify when gas leaves the CGM. 
For example, gas that approaches the central galaxy but is blown away by a galactic wind before being incorporated the galaxy can momentarily cross the (relatively large) $R_{\rm gal}$ radius. 
To avoid classifying such gas as accreted by the central galaxy, we only consider gas to have properly left the CGM if it is not contained by the galaxy-halo interface. 
Another means through which gas may spuriously appear to leave the CGM is through the momentary misidentification of the main halo by the halo finder. 
This causes the position of the main halo's virial radius to change abruptly and can result in spuriously ``ejected' gas.
To minimize this, we require that gas particles spend $t > 30$ Myr in a destination component to count as having left the CGM. 
Since the spacing between simulation snapshots is $\sim 25$ Myr, this  requires that gas particles spend $\ge 2$ snapshots in the destination component (there are some more finely spaced snapshots at very high redshift).

\section{Results}
\label{sec:results}

\subsection{Past and Future Locations of CGM Gas}
\label{sec:baryon_cycle}

\begin{figure*}
\centering
\begin{minipage}{0.495\textwidth}
\centering
\includegraphics[width=\textwidth]{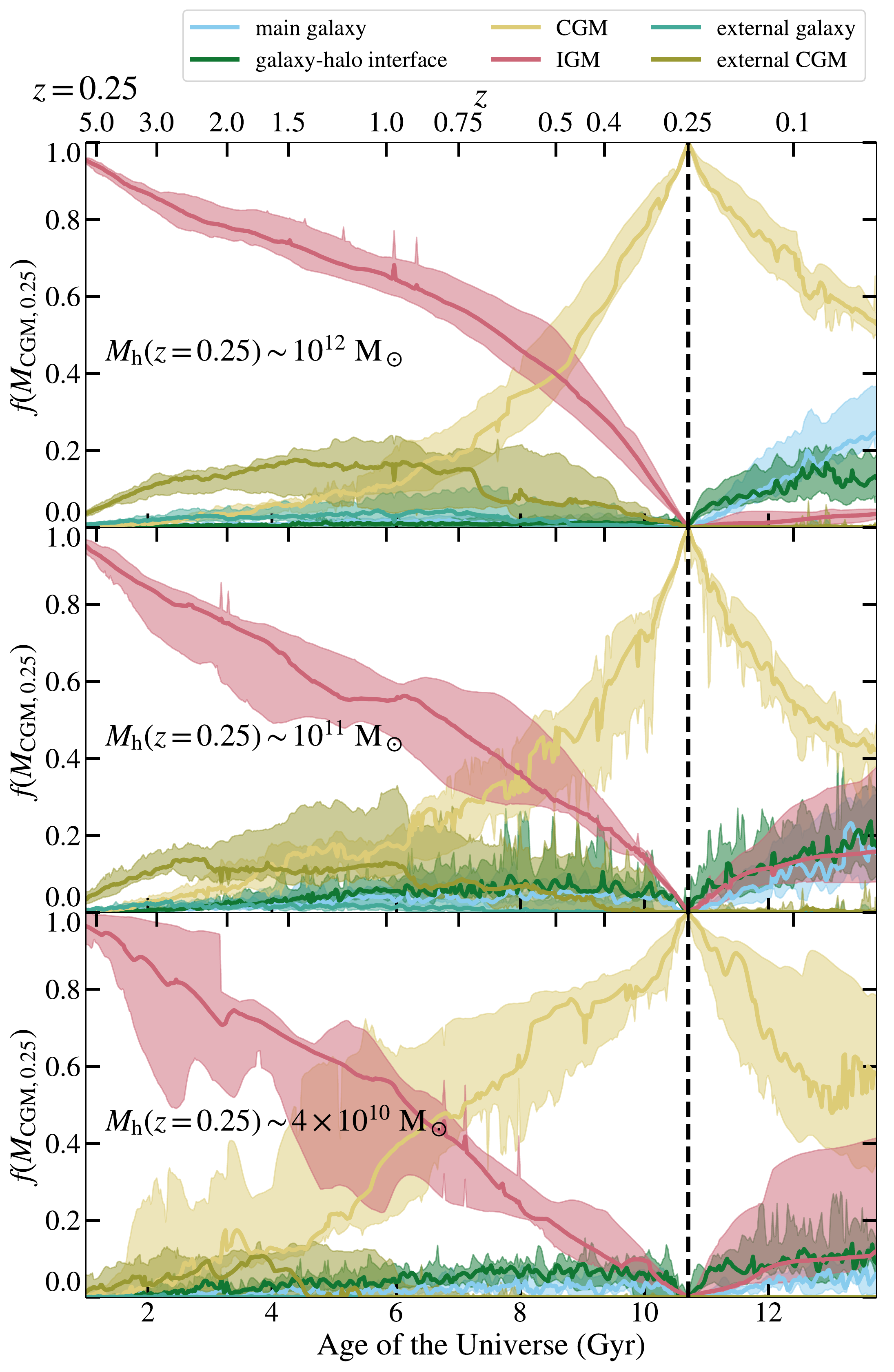}
\end{minipage} \hfill
\begin{minipage}{0.495\textwidth}
\centering
\includegraphics[width=\textwidth]{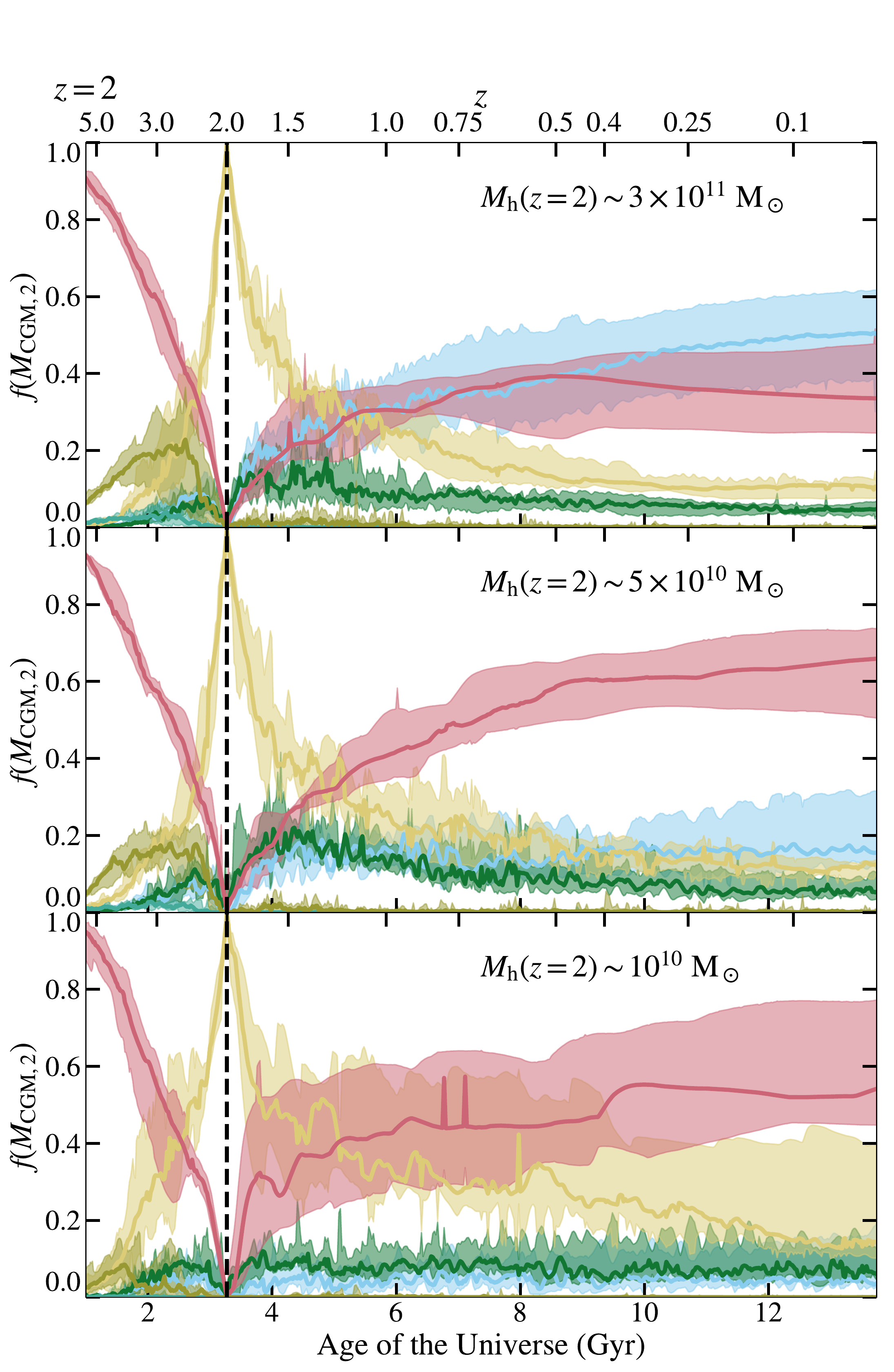}
\end{minipage} \hfill
\caption{
Fraction of mass in a given component of the baryon cycle as a function of time for gas that was/will be in the CGM at $z=0.25$ \textbf{(left)} and $z=2$ \textbf{(right)}. 
We show gas in the main galaxy (blue),
the galaxy-halo interface of the main galaxy (green),
the CGM of the main galaxy (yellow),
the IGM outside $R_{\rm vir}$ of any galaxy (red),
within a galaxy other than the main galaxy (teal),
and within the CGM of a galaxy other than the main galaxy (dark yellow). 
The solid line shows the median across different halos in a given mass bin, and the shaded regions indicate the 16th to 84th percentiles between halos. 
Across the halo mass range analyzed, both the build-up and the subsequent loss of half of the CGM mass occurs over $\sim 3$ Gyr for $z=0.25$ CGM gas and $\lesssim 1$ Gyr for $z=2$ CGM gas.
}
\label{fig:CGM_gas_location}
\end{figure*}

Figure~\ref{fig:CGM_gas_location} shows the results of identifying gas in the CGM at $z=0.25$ or $z=2$, and tracking the fraction of that mass residing in different components of the baryon cycle at subsequent and prior redshifts.
The fractions at $z=0$ correspond to ultimate fates, which we also study as a function of halo mass in  \S\ref{sec:fate_by_halo_mass}. 

As mentioned in the introduction, several CGM surveys in the last decade have targeted $z\sim0.25$ using the Cosmic Origins Spectrograph on HST or $z\sim2$ using 10m-class ground-based observatories.
As indicated by the yellow curves in the left panels of Figure~\ref{fig:CGM_gas_location}, the fraction of $z=0.25$ CGM gas that was in the CGM of the same galaxy at $z=2$ is on average $\lesssim 10\%$ for all halo masses included in our analysis.
Thus, our simulations indicate that CGM surveys at different redshifts primarily probe different gas. 

Of the CGM at $z=2$, about half ends up in the central galaxy by $z=0$ in $10^{12} \Msun$ progenitor halos; of the rest most is ejected into the IGM. 
Most of the CGM of lower-mass halos at $z=2$ is ejected into the IGM over time. 
Interestingly, this is so even though most of the CGM in $M_{\rm h}(z=2) \sim 5\times10^{10}$ M$_{\odot}$ halos first accretes onto the central galaxy before being ejected into the IGM, as implied by our analysis of proximate fates in \S\ref{sec:fate_by_halo_mass}. 
This highlights the complex dynamics of inflows and outflows in the CGM, and in particular the fact that much of the gas accreted by galaxies can later be ejected in powerful winds, especially in low-mass halos \citep[e.g.,][]{Angles-Alcazar2017}. 
On the other hand, a significant fraction of the CGM mass at $z=0.25$ remains in the CGM by $z=0$ at all halo masses analyzed.

Up to $\sim 20\%$ of the CGM mass at a given redshift spends time in halos other than the main halo prior to accreting onto the main CGM. 
Of this mass, the vast majority is in the CGM of other halos (as opposed to in external galaxies; see the dark yellow curves in Figure~\ref{fig:CGM_gas_location}).
This external gas ends up in the main halo either when the external halo is accreted onto the main halo 
or by being expelled by the external halo and subsequently accreting onto the CGM of the main galaxy (see the bottom panel of Figure 8 of \originp~for an example of this for a $10^{10} \Msun$ progenitor). 
This latter channel is related to the ``intergalactic transfer'' identified in FIRE-1 simulations by \cite{Angles-Alcazar2017}, but in this case consists primarily of CGM-to-CGM transfer, rather than galaxy-to-galaxy transfer.

\subsection{Pathlines for Different CGM Fates}
\label{sec:particle_pathlines}

\begin{figure*}
\centering
\includegraphics[height=0.29\textheight]{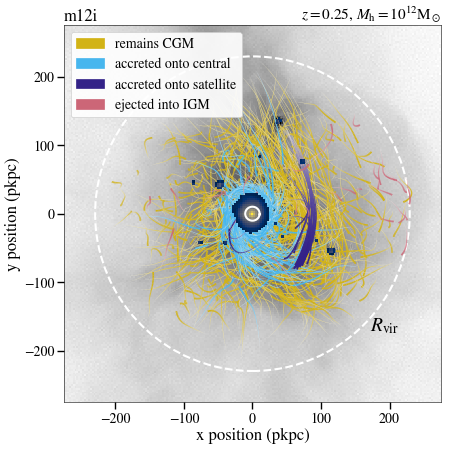}
\includegraphics[height=0.29\textheight]{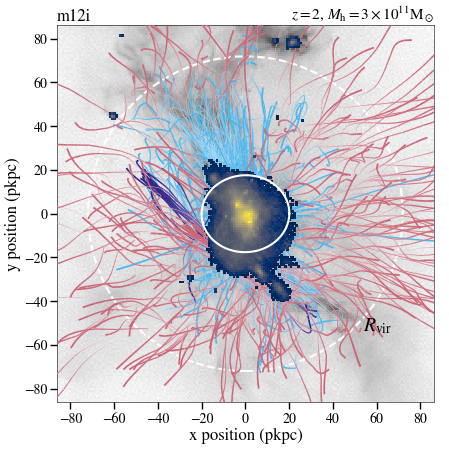}
\includegraphics[height=0.29\textheight]{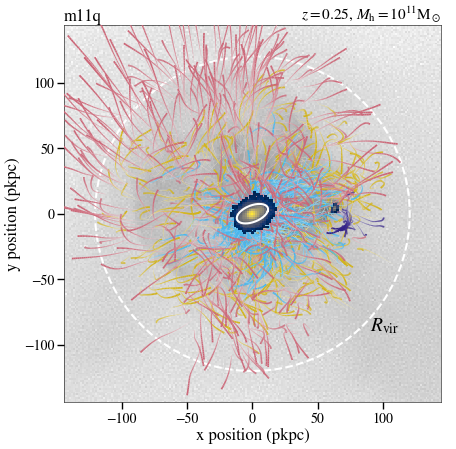}
\includegraphics[height=0.29\textheight]{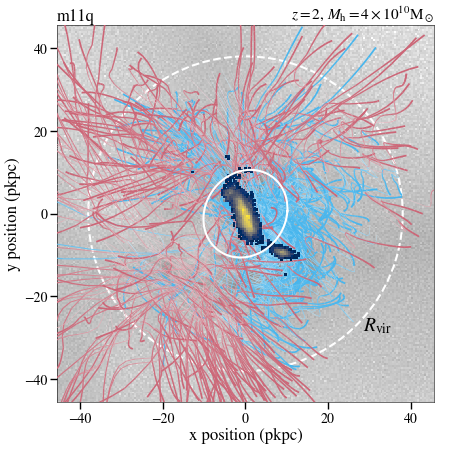}
\includegraphics[height=0.29\textheight]{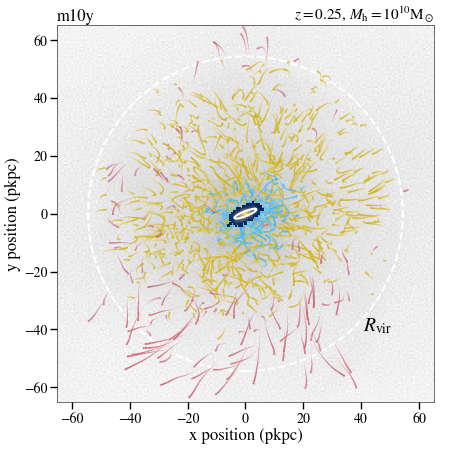}
\includegraphics[height=0.29\textheight]{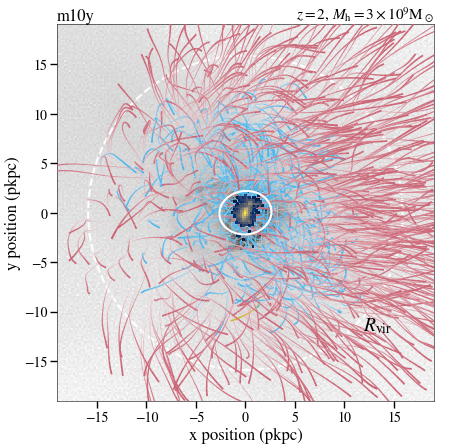}
\caption{
\textbf{Left:} Paths traced over the course of 1 Gyr by 1000 randomly-selected particles in the CGM of \texttt{m12i} (top), \texttt{m11q} (center), and \texttt{m10y} (bottom).
Pathlines start at $z=0.25$ and are colored according proximate fate (i.e., the fate of gas immediately after it leaves the CGM). 
Lines become thicker with increasing time. 
The stellar mass surface density is plotted as a blue-yellow histogram. 
\textbf{Right:} Same as left, but for $z=2$.
To account for shorter halo dynamical times at this cosmic time, particle trajectories are plotted over the course of 0.5 Gyr, as opposed to 1 Gyr. 
In each panel, the virial radius of the main galaxy is plotted as a dashed white circle, and a circle with radius $R_{\rm gal}$ is centered on the main galaxies.
The galaxy circle is rotated such that its plane is normal to the total angular momentum of the galaxy's stars, indicating the orientation of the galactic disk. 
At $z=2$ in the $\gtrsim 10^{11} \Msun$ progenitors (\texttt{m12i} and \texttt{m11q}) gas that will be ejected into the IGM is often spatially separated from gas that will accrete onto the central galaxy, indicating that accreting gas can shape the paths of outflows.
}
\label{fig:pathlines_CGM}
\end{figure*}

\begin{figure}
\includegraphics[width=\columnwidth]{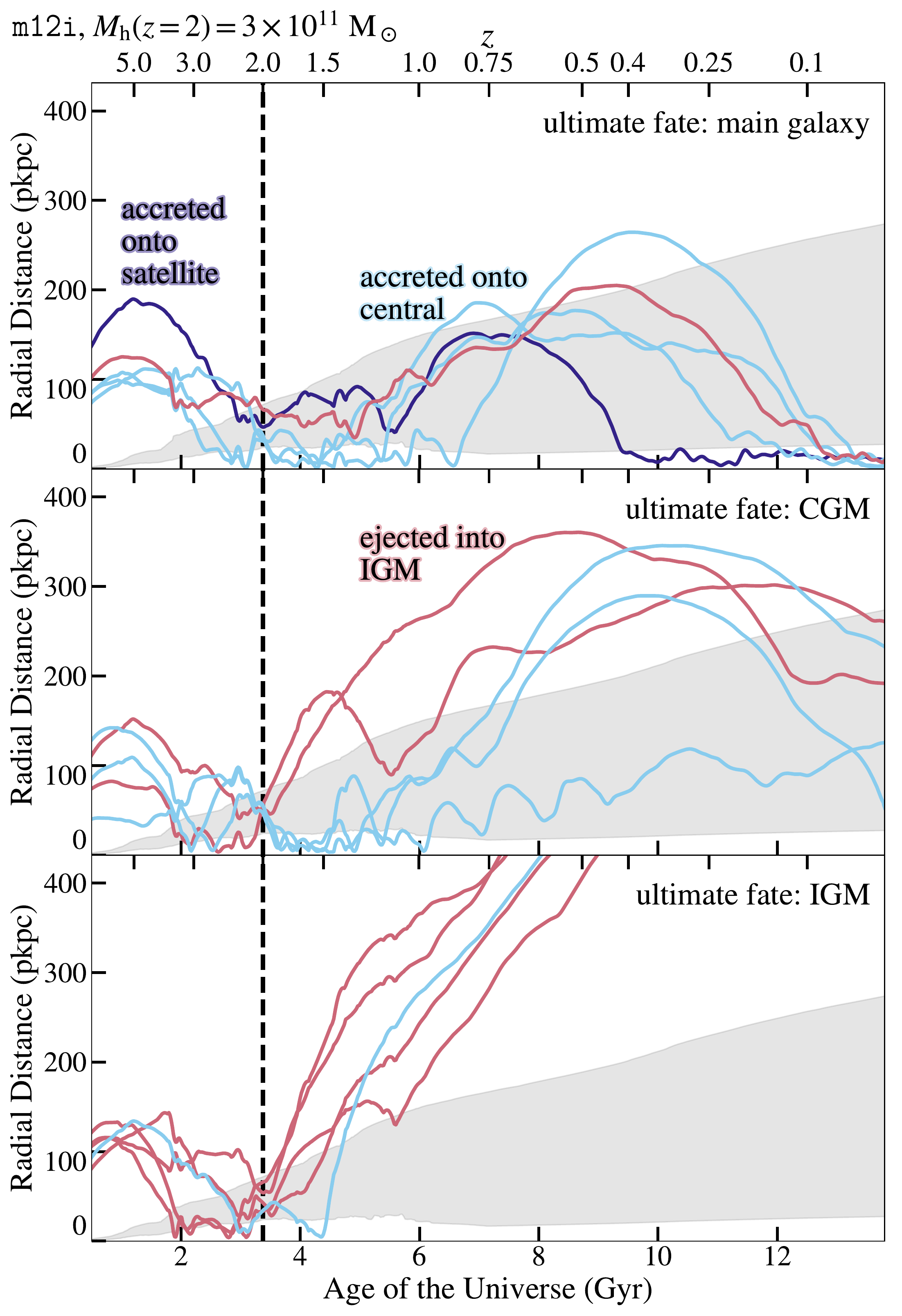}
\caption{
Distance from the central galaxy (in proper kpc) vs age of the universe for particles located in the CGM of \texttt{m12i} at $z=2$, a halo with  $M_{\rm h}(z=2) = 3 \times 10^{11} \Msun$.
The grey shaded region shows the boundaries of the CGM.
A vertical dashed line displays $z=2$.
Each panel corresponds to the location of the gas at $z=0$ (its ultimate fate), with five particles per panel.
Gas that will eventually be part of the central galaxy, the CGM of the main halo, and the IGM are the top, middle, and bottom panels respectively.
The color of the gas particle indicates its proximate fate: gas accreted onto the main galaxy, onto satellites, or ejected into the IGM is colored blue, purple, and red, respectively.
Gas with a given ultimate fate can reach its destination through a wide variety of possible trajectories.
}
\label{fig:r_vs_time_eventual_m12i_CGM_snum172}
\end{figure}

In this section we analyze how the trajectories of particles found in the CGM of main halos depend on their proximate and ultimate fates. 
The visualizations allow us to better understand the spatial distributions and physical behaviors of gas with different fates, and to illustrate how sight lines through halos will in general include gas of different fates.

Figure~\ref{fig:pathlines_CGM} shows the paths traced by 1000 randomly selected gas particles in the CGM of three representative simulations, one for each of our main halo mass bins. 
On the left, the paths are shown for 1 Gyr after $z=0.25$ and on the right, for 0.5 Gyr after $z=2$.
The number of particles displayed for each proximate fate is proportional to the mass contribution of that fate to the CGM.
\footnote{
Visualizations of the same simulations were presented for CGM origins in \originp.
However, in that study 500 gas particles were selected \emph{for each origin} (independent of the origin's mass contribution CGM). 
The visualizations shown in the present paper thus more faithfully represent the contributions of different processes to the total CGM mass.
}

Gas that will be accreted onto satellite galaxies closely follows the trajectory of satellite galaxies.
Gas accreted onto satellites can be subsequently ejected from satellite galaxies.
As apparent in some of the panels of Figure~\ref{fig:pathlines_CGM} (e.g. for \texttt{m12i} at $z=0$), some of the gas diverges from satellites after having been accreted by a satellite galaxy, indicating an outflow from the satellite.
We discuss this more in \S\ref{sec:origin_fate_connection}, where we show that the primary origin of gas whose proximate fate is to be accreted onto a satellite is satellite wind. 
This reflects that fact that (relatively low-mass) star-forming satellite galaxies generally drive multiple episodes of gas ejection and recycling \citep[see also][]{Angles-Alcazar2017}. 

As expected, gas whose proximate fate is to be accreted onto the central galaxy is preferentially found close to the central galaxy. 
Gas that will be ejected into the IGM appears to have quite diverse spatial distributions. 
In some cases gas that will accrete onto the central galaxy appears to block outflowing gas (e.g. \texttt{m12i} and \texttt{m11q} at $z=2$). 
Interestingly, the $M_{\rm h}(z=0) \sim 10^{12}$ M$_{\odot}$ halo shown in Figure \ref{fig:pathlines_CGM} contains very little gas at $z=0.25$ that will be ejected into the IGM. 
This is consistent with the general finding in the FIRE simulations that galactic winds essentially disappear at low redshift in $\sim L^{\star}$ galaxies~(e.g. \citealt{Muratov2015, Angles-Alcazar2017}, Stern et al. in prep.).

Figure~\ref{fig:r_vs_time_eventual_m12i_CGM_snum172} shows radial distance from the center of the halo versus time for particles found in the CGM at $z=2$ in our example $10^{12} \Msun$ progenitor, \texttt{m12i}.
Equivalent plots for characteristic $10^{11}$ and $10^{10} \Msun$ progenitors are found in Appendix~\ref{sec:more_pathlines}.
We visualize 5 particles in each ultimate fate panel.  
Within each panel, the lines are colored according to proximate fate.
The particles visualized here were selected to illustrate the different proximate fates and wide variety of trajectories possible for a given ultimate fate. 
Consistent with the bursty outflows in FIRE, gas that ends up in the central galaxy frequently recycles across all halo masses included in our analysis. 

\subsection{Fates of the CGM}
\label{sec:fates_of_the_CGM}

\subsubsection{Fate by Halo Mass}
\label{sec:fate_by_halo_mass}

\begin{figure*}
\centering
\includegraphics[width=0.9775\linewidth]{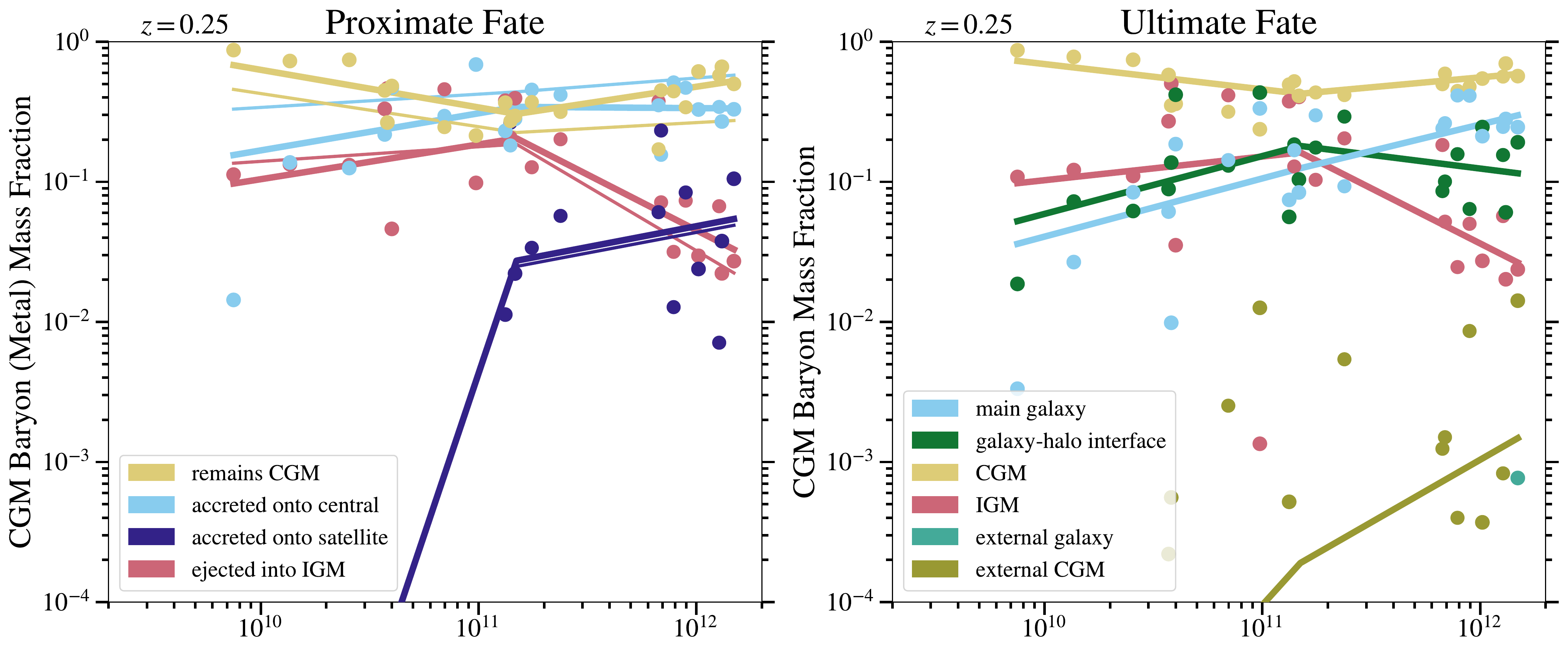}
\includegraphics[width=\linewidth]{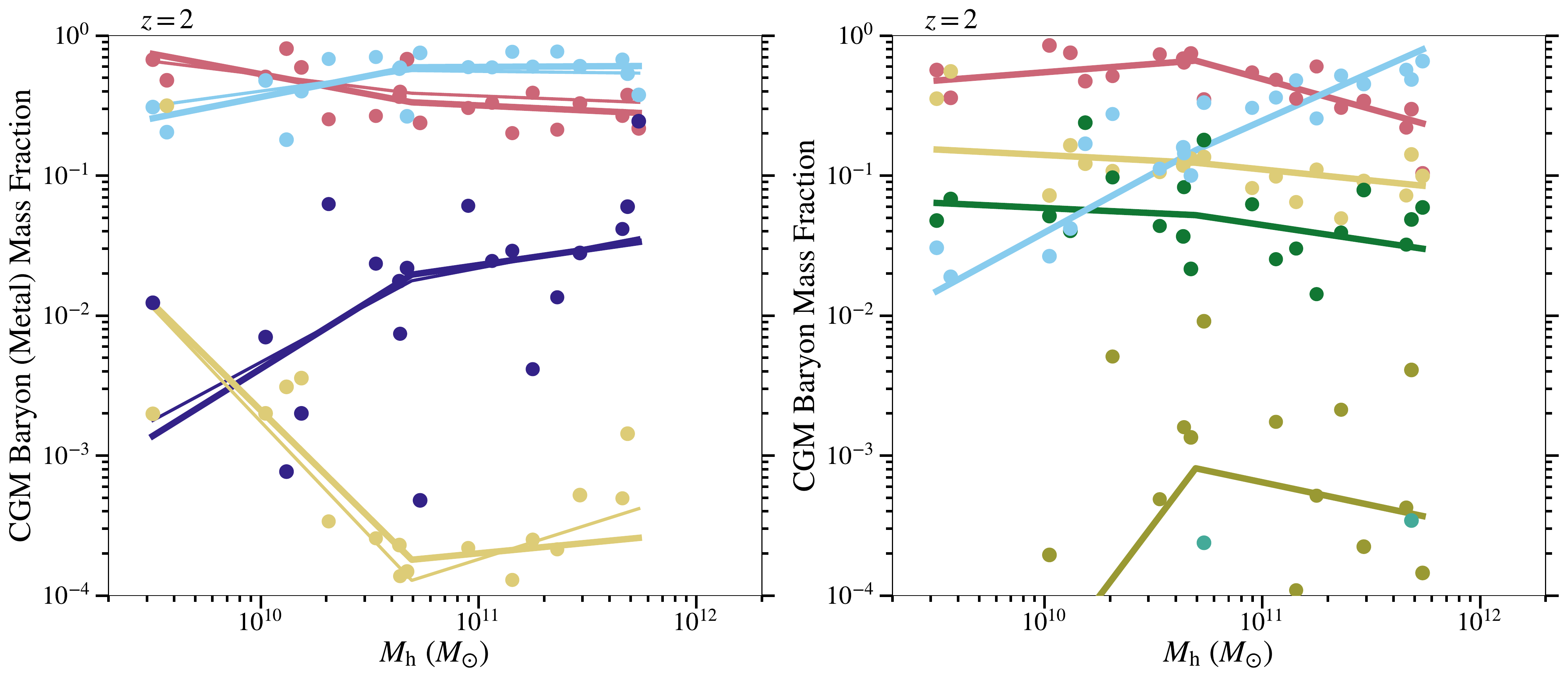}
\caption{
\textbf{Left:}
Fraction of the total CGM baryon mass at $z=0.25$ (top) or $z=2$ (bottom) for different proximate fates (i.e. the component of the baryon cycle the gas transfers to upon leaving the CGM). 
Each point is a value from a simulation.
The thick lines connect the medians for each mass bin. 
The thin lines are calculated similarly, but for the fraction of CGM \textit{metal} mass with different proximate fates, for which the raw data is not shown here.
\textbf{Right:} Similar but for fractions with different ultimate fates
Similar but for the fraction of the total CGM baryon mass with different ultimate fates, i.e. the component of the baryon cycle where the gas end up by $z=0$.
}
\label{fig:CGM_mass_frac_vs_Mh_CGM}
\end{figure*}

The left side of Figure~\ref{fig:CGM_mass_frac_vs_Mh_CGM} shows the fraction of CGM mass with different proximate fates at $z=0.25$ and $z=2$.
The right side shows the same but for different ultimate fates.
The points are for individual halos and the lines connect the medians between each halo mass bin.
We highlight below some important trends regarding the possible fates of the CGM, as well conclusions that can be drawn by comparing proximate and ultimate fates. 

One evident difference in Figure~\ref{fig:CGM_mass_frac_vs_Mh_CGM} between $z=0.25$ and $z=2$ is that while, overall, the most common proximate fate for CGM gas at $z=0.25$ is to remain in the CGM until $z=0$, only a tiny fraction $\sim 10^{-4}-10^{-2}$ of the $z=2$ CGM mass remains continuously in the CGM until the present time. 
Several factors contribute to this difference:
(i) there is more time for gas to leave the CGM between $z=2$ and $z=0$;
(ii) characteristic halo dynamical timescales are shorter at $z=2$; and
(iii) in the case of the more massive $M_{\rm h}(z=0)\sim10^{12}$ M$_{\odot}$ progenitors at $z=0.25$, the CGM forms a hot, pressure-supported atmosphere where accretion occurs on a cooling time, which is longer than the dynamical time~\citep{Hafen2019a, Stern2019, Stern2019a}. 
The very small ``remains CGM'' proximate fate fractions at $z=2$ however do \emph{not} imply there is essentially no gas in common between the CGM of galaxies at $z=2$ and $z\sim0$. 
Indeed, Figure~\ref{fig:CGM_mass_frac_vs_Mh_CGM} shows that $\sim10$\% of the $z=2$ CGM is also in the CGM of the main galaxy at $z=0$. To reconcile these results, we conclude that while almost all the $z=2$ CGM leaves the CGM at least once by $z=0$, a significant fraction \emph{returns} to the CGM by $z=0$. 
Gas accreted by galaxies can return to the CGM via galactic winds, while gas ejected into the IGM can later fall back onto the CGM.

Accretion onto the central galaxy is a common proximate fate for much of the CGM across the halo mass and redshift ranges analyzed. On average, the prevalence of this proximate fate increases with halo mass and redshift. 
For $M_{\rm h}(z=0)\sim10^{12}$ M$_{\odot}$ progenitors at $z=2$, corresponding to observed Lyman break galaxies \citep[LBGs; e.g.,][albeit at the low-mass end]{Adelberger2005}, $\gtrsim 50\%$ of the CGM mass has this proximate fate. 
Moreover, the simulations indicate that up to $\sim 80\%$ of the CGM of LBGs end up in central galaxies by $z=0$ (ultimate fate). 
The increasing fraction of the CGM mass that ends up in central galaxies (at both $z=0.25$ and $z=2$) is in agreement with the increasing efficiency with which observations imply that halos convert their baryon budget into stars \citep[e.g.,][]{Moster2013a, Behroozi2013}. 

The thin lines in the left column of Figure~\ref{fig:CGM_mass_frac_vs_Mh_CGM} show CGM metal mass fractions with different proximate fates.
The most significant differences between the proximate fates of the overall CGM mass vs. the CGM metals are in the fractions of the mass/metals that will remain in the CGM vs. will be accreted onto the central galaxy. 
In particular, the figure shows that metal-enriched gas is preferentially accreted relative to the overall CGM mass. 
In our sample, the vast majority of CGM metals originate in galactic winds (see \originp, Figure 10). 
The results thus imply that, at $z=0.25$, CGM mass previously contributed by winds is more likely to reaccrete onto the central galaxy than other CGM mass (e.g., from IGM accretion). 
This effect is due both to the fact that the CGM mass fraction from previous winds is enhanced in the inner halo (Figure 14 in \originp) and to enhanced cooling of metal-rich gas (which tends to be overdense in addition to metal-rich; Esmerian et al., in prep.). 
Figure~\ref{fig:CGM_mass_frac_vs_Mh_CGM} shows that at $z=2$ metallicity does not affect the relative fraction of accreting total gas mass vs. metals, which is due in at least part to the almost complete absence of $z=2$ CGM gas that remains continuously in the CGM until $z=0$.

As discussed in \S \ref{sec:simulations}, a subset of our simulations include a subgrid prescription for metal diffusion that allows metals to diffuse between adjacent resolution elements.\footnote{Technically, a metallicity scalar field is diffused but there is no actual transfer of mass between MFM resolution elements.}
We use both simulations produced with and without subgrid metal diffusion to calculate the fractions 
in Figure~\ref{fig:CGM_mass_frac_vs_Mh_CGM}.
This effectively assumes that the fates of CGM mass and metals are not significantly affected by subgrid metal diffusion. 
To check this assumption we calculated the CGM fractions with different proximate fates separately for simulations with and without metal diffusion, and found no differences that could not be explained by halo-to-halo variations and the small number of halos.
However we caution that our simulation sample is relatively small, so significant differences could be revealed in a larger sample. 

Overall, the accretion of CGM gas onto satellite galaxies is a subdominant proximate fate.
However, this process becomes increasingly important with increasing halo mass, at both $z=0.25$ and $z=2$, and can be the proximate fate of up to $\sim 10\%$ of the CGM mass in $10^{12} \Msun$ halos.

Concerning the ejection of gas into the IGM, Figure~\ref{fig:CGM_mass_frac_vs_Mh_CGM} shows that this is the dominant ultimate fate of the $z=2$ CGM of most the halos included in our analysis -- the exception being the more massive $M_{\rm h}(z=0)\sim10^{12}$ M$_{\odot}$ progenitors, for which the simulations instead predict that most of the CGM ends up in the central galaxy. 
Ejection into the IGM is a dominant fate even though the proximate fate results show that most of the CGM in $M_{\rm h}(z=2) \sim 5\times10^{10} \Msun$ halos first accretes onto the central galaxy before being ejected into the IGM. 
This difference between proximate and ultimate fates again highlights the complex dynamics of the baryonc cycle. 

\subsubsection{Fates by Temperature}
\label{sec:temperature}

\begin{figure*}
\centering
\begin{minipage}{0.495\textwidth}
\centering
\includegraphics[width=\textwidth]{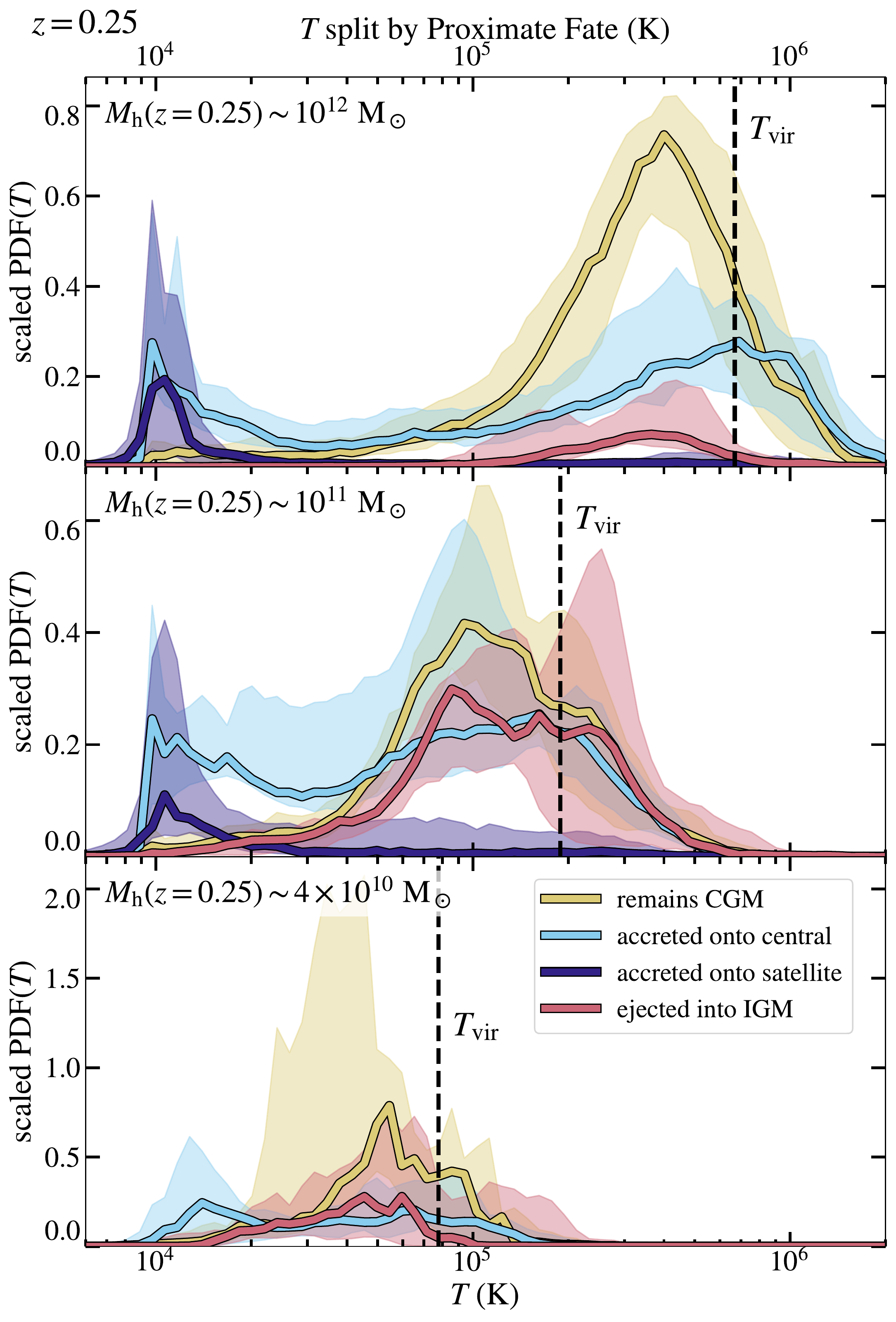}
\end{minipage} \hfill
\begin{minipage}{0.495\textwidth}
\centering
\includegraphics[width=\textwidth]{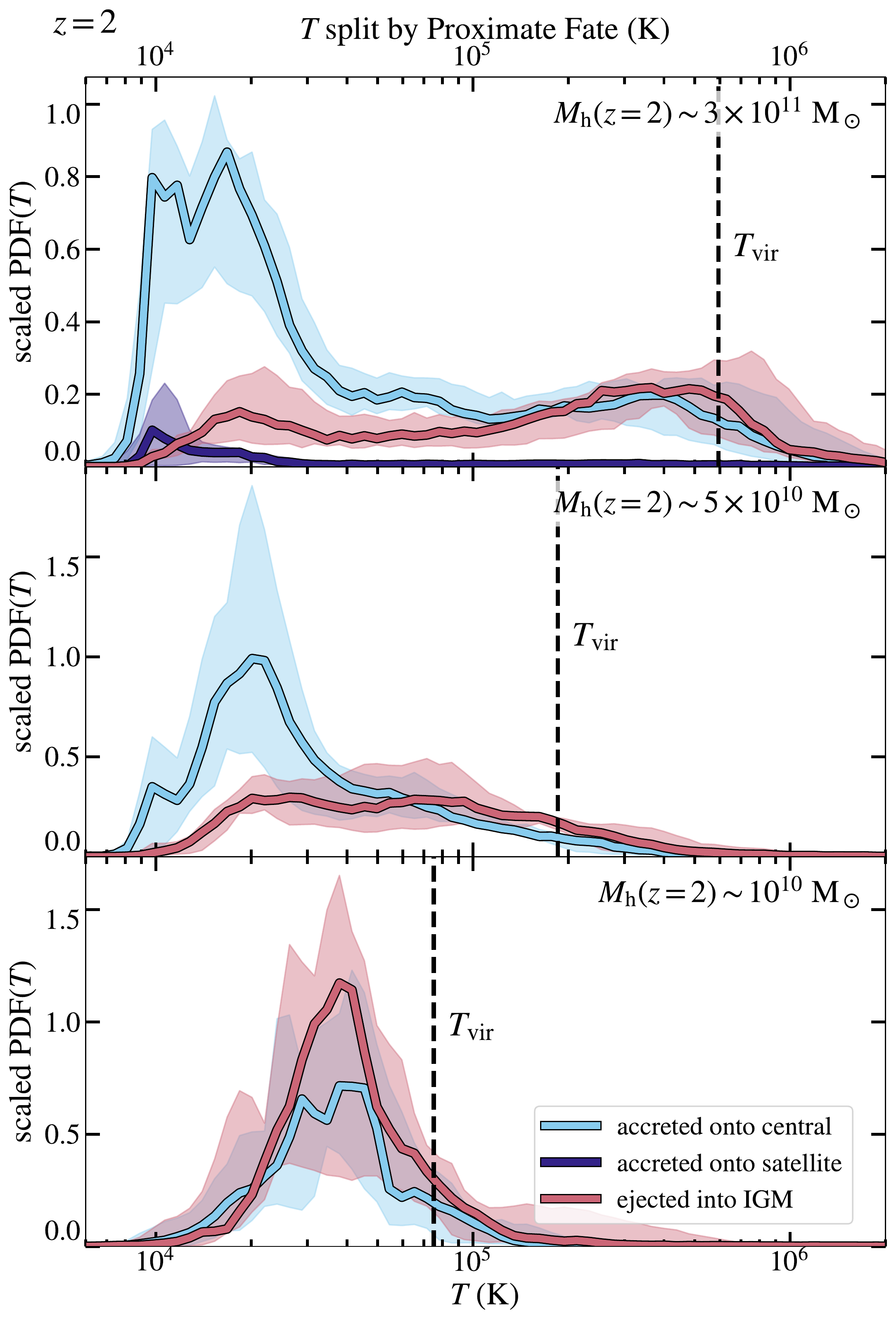}
\end{minipage} \hfill
\caption{
The temperature distribution of CGM gas for different proximate fates at $z=0.25$ \textbf{(left)} and $z=2$ \textbf{(right)}. In each panel, the temperature distributions are weighted by the contributions of the different fates to the total CGM mass.
Top, middle, and bottom panels correspond to halos with $M_{\rm h} (z=0) \sim 10^{(12,11,10)} \Msun$, respectively.
Solid lines indicate the median value for the temperature bin across different halos, while the shaded region encompasses the 16th-to-84th percentiles.
The vertical dashed lines indicate halo virial temperatures.
}
\label{fig:CGM_temp}
\end{figure*}

Figure~\ref{fig:CGM_temp} shows the temperature distributions of CGM gas of different proximate fates, for different halo mass bins. 
The solid lines indicate the median between halos in a given halo mass bin, and the shaded regions indicate the 16th-to-84th percentile range.
For each proximate fate, the $\log T$ probability distribution function (PDF) is weighted by the fraction of mass contributed to the CGM.
Thus, an integral over $\log T$ corresponds to the fraction of the total CGM mass contributed by a given proximate fate.
At a given temperature the relative heights of the PDFs indicate the relative contribution of different proximate fates to CGM gas mass in a $\Delta \log{T}$ temperature interval.

One striking result in Figure \ref{fig:CGM_temp} is that, at $z=2$, most of the CGM in the halos analyzed is cool, with $T \ll T_{\rm vir}$ ($T_{\rm vir} =  G M_{\rm h} \mu m_{\rm p} / ( 2 k_{\rm B} R_{\rm vir} )$ is the halo virial temperature, where $\mu=0.6$ is the assumed mean molecular weight, $m_{\rm p}$ is the proton mass, and $k_{\rm B}$ is the Boltzmann constant).
Of this cool gas, most of it will next accrete onto the central galaxy in the progenitors of $\sim 10^{11}-10^{12}$ M$_{\odot}$ halos. 
This is consistent with the importance of cold mode galactic accretion for these high-redshift halos \citep[e.g.,][]{Birnboim2003, Keres2005, Keres2009}. 
On the other hand, for the $M_{\rm h}\sim10^{12}$ M$_{\odot}$ halos at $z=0.25$, we find that most of the CGM is hot with $T \sim T_{\rm vir}$ and remains in the CGM until $z=0$.
Of the gas that accretes onto the central galaxy by $z=0$, most of it is hot at $z=0.25$, consistent with the development of hot mode accretion in the halos of low-redshift $\sim L^{\star}$ galaxies. 

Focusing on the fate of cool gas with $T \approx 10^4$ K, Figure \ref{fig:CGM_temp} implies that (to the extent that the present simulations are realistic), most cool gas observed in the CGM of halos in the parameter space covered by our analysis is likely to next accrete onto a galaxy (except perhaps for the low-mass progenitors of $\sim 10^{10}$ M$_{\odot}$ at $z=2$, which eject a lot of mass into the IGM). 
Interestingly, for $10^{12} \Msun$ halos at $z=0.25$, about half of the cool CGM with $T \approx 10^4$ K will next accrete onto satellite galaxies. 
As we will discuss further in \S \ref{sec:origin_fate_connection}, this is because cool gas in hot halos in our simulations is strongly associated with non-linear structures, in particular satellite winds that recycle.
This is especially so at larger halo radii, where satellite wind is more prevalent than wind from the central galaxy (\originp, Figure 14). 

\subsubsection{Fates by Metallicity}
\label{sec:metallicity}

\begin{figure*}
\centering
\begin{minipage}{0.495\textwidth}
\centering
\includegraphics[width=\textwidth]{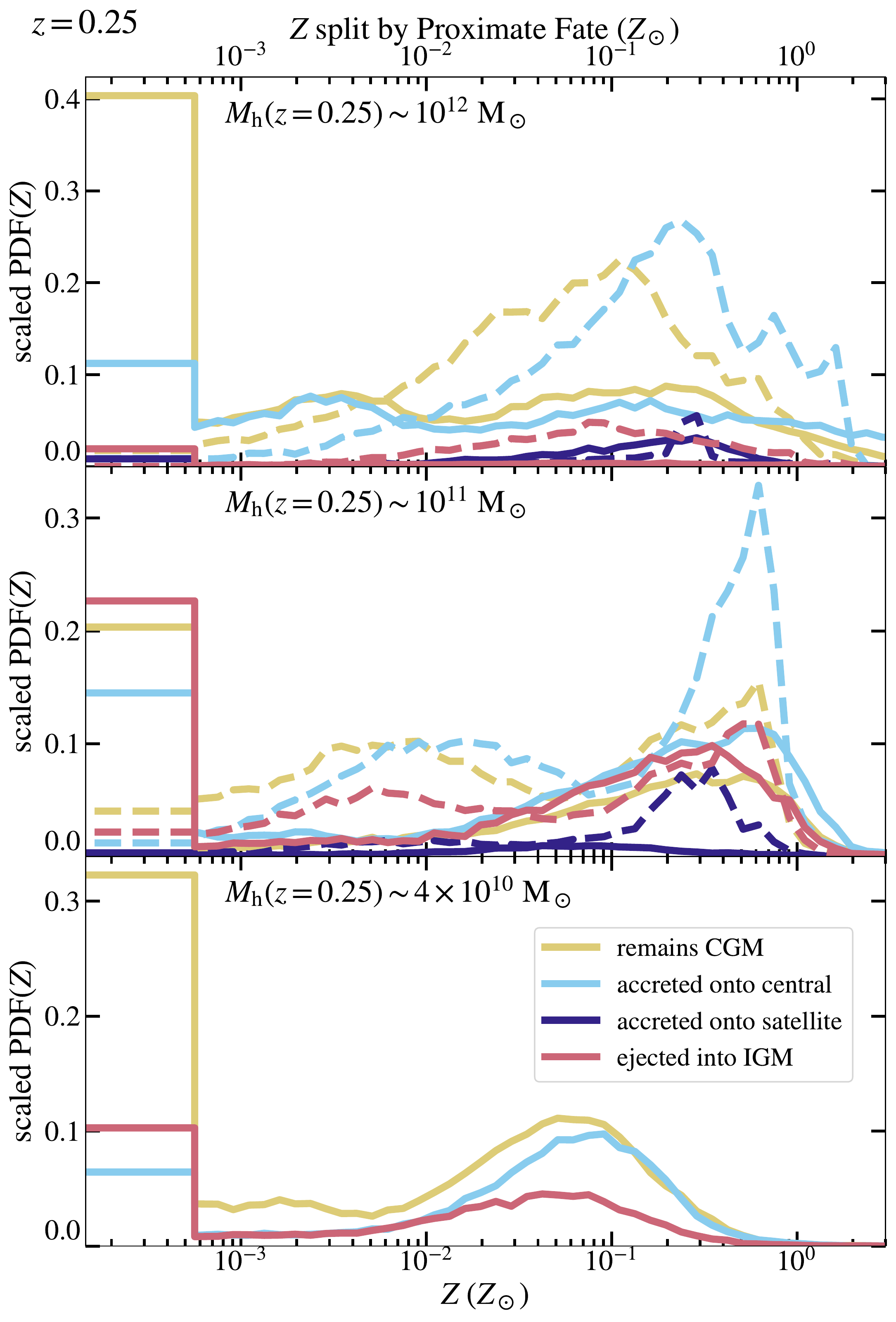}
\end{minipage} \hfill
\begin{minipage}{0.495\textwidth}
\centering
\includegraphics[width=\textwidth]{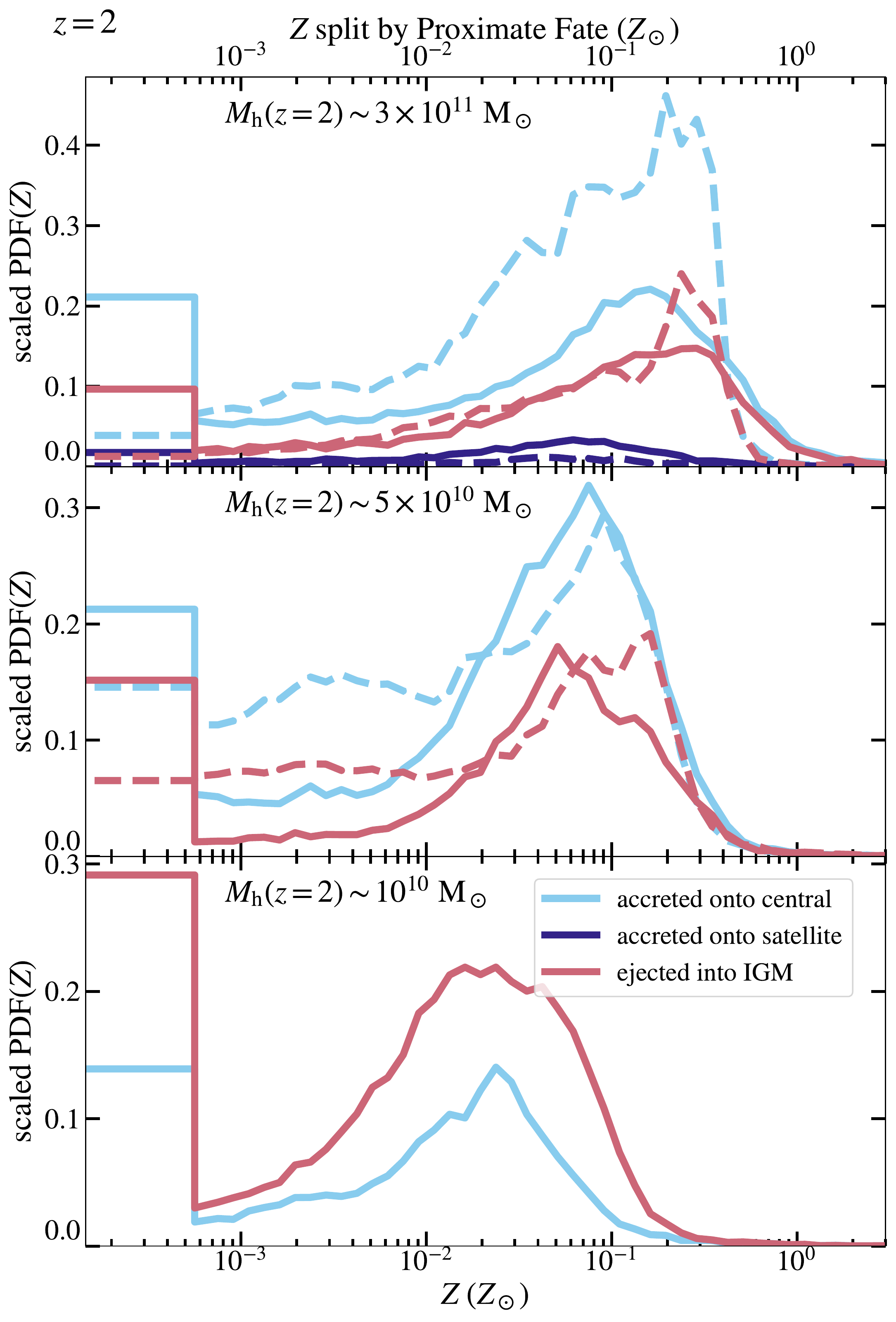}
\end{minipage} \hfill
\caption{
The metallicity distribution of CGM gas with a given proximate fate weighted by the contribution of such gas to the total CGM mass.
The solid (dashed) lines indicate the median value of the distribution across all simulations in that mass bin run with (without) a prescription for subgrid turbulent diffusion.
While the distributions change for simulations with/without metal diffusion, for both types of simulations and at both $z=0.25$ and $z=2$ the metallicity distributions of different fates overlap strongly, indicating that the metallicity of CGM gas elements is not a reliable indicator of its fate. 
While the distributions overlap strongly, at $z=0.25$ the median metallicity of gas accreted onto galaxies can be up to $\sim 1$ dex higher than the median metallicity of gas that will remain in the CGM or be ejected into the IGM by $z=0$.
}
\label{fig:CGM_metallicity}
\end{figure*}

As demonstrated explicitly in \originp, the metallicity of CGM gas is a strong function of its origin (i.e. as galactic winds vs. IGM accretion) and the host halo mass.
In this section we investigate the connection between metallicity and the proximate fate of CGM gas.
Figure~\ref{fig:CGM_metallicity} shows the metallicity distributions of CGM gas, with solid (dashed) lines indicating the median between halos in a given halo mass bin for simulations run without (with) turbulent metal diffusion. 
The distributions are normalized by contribution to the total CGM mass, as we normalized the temperature distributions in the previous section.
We use the same value of the solar metallicity as used in \originp, $Z_\odot = 0.0134$~\citep{Asplund2010}.
Note that our simulations have a metallicity floor at $Z \approx 10^{-4} Z_\odot$.
To account for a spike in the values of the PDF near the metallicity floor we expand the bottom-most histogram bin to cover $Z = 10^{-4} - 10^{-3.5} Z_\odot$.

Figure~\ref{fig:CGM_metallicity} shows that, for a given halo mass bin and redshift, the CGM metallicity distributions of different proximate fates are broad and overlap substantially. 
As a result, the simulations indicate that we cannot in general use the metallicity of a CGM absorber to reliably predict whether the gas will next accrete onto galaxy, be ejected into the IGM, or remain in the CGM. 
We note, however, that while metallicity distributions of different proximate fates overlap broadly, the median metallicity for different proximate fates can differ significantly.
For example, at $z=0.25$ the median metallicity of gas that will accrete onto galaxies is $\sim 1$ dex higher than the metallicity of gas that will remain in the CGM until $z=0$. 

The results for simulations including a subgrid model turbulent metal diffusion (see \S \ref{sec:simulations}), are shown by dashed lines in Figure \ref{fig:CGM_metallicity}. 
In these simulations, low-metallicity gas particles gain metals through contact with higher metallicity gas. 
Subgrid metal diffusion can significantly boost the metallicity of otherwise metal-poor gas. 
As can be seen in Figure \ref{fig:CGM_metallicity}, this almost entirely suppresses the metallicity floor peaks for the CGM of all possible fates. 
However, since metallicity distributions for different proximate fates nevertheless overlap strongly, subgrid metal diffusion does not affect our qualitative conclusion that metallicity is not in general a robust predictor of the fate of a CGM gas element.

The inner and outer CGM typically probe very different physical regimes:
wind from the central galaxy can provide up to $\sim 80\%$ of the mass in the inner CGM while IGM accretion can provide a similar fraction of the CGM mass at $R \sim R_{\rm vir}$~(\originp).
We analyzed the metallicity distributions of proximate fates as a function of radius and found strong overlap between the metallicity distributions at all radii, qualitatively similar to the results in Figure~\ref{fig:CGM_metallicity}.

\subsection{Connection Between Origin and Fate}
\label{sec:origin_fate_connection}

\begin{figure*}
\centering
\begin{minipage}{0.495\textwidth}
\centering
\includegraphics[width=\textwidth]{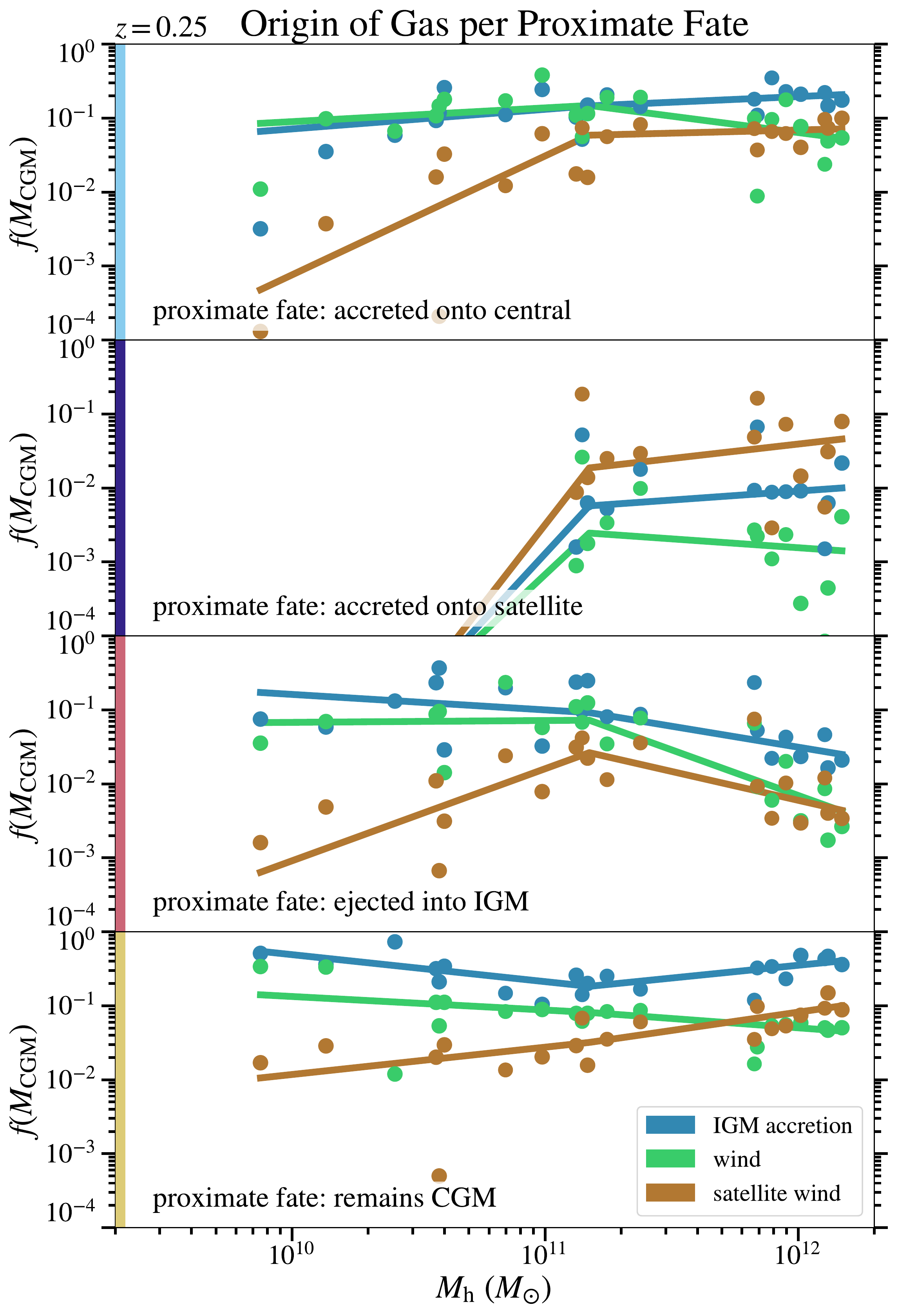}
\end{minipage} \hfill
\begin{minipage}{0.495\textwidth}
\centering
\includegraphics[width=\textwidth]{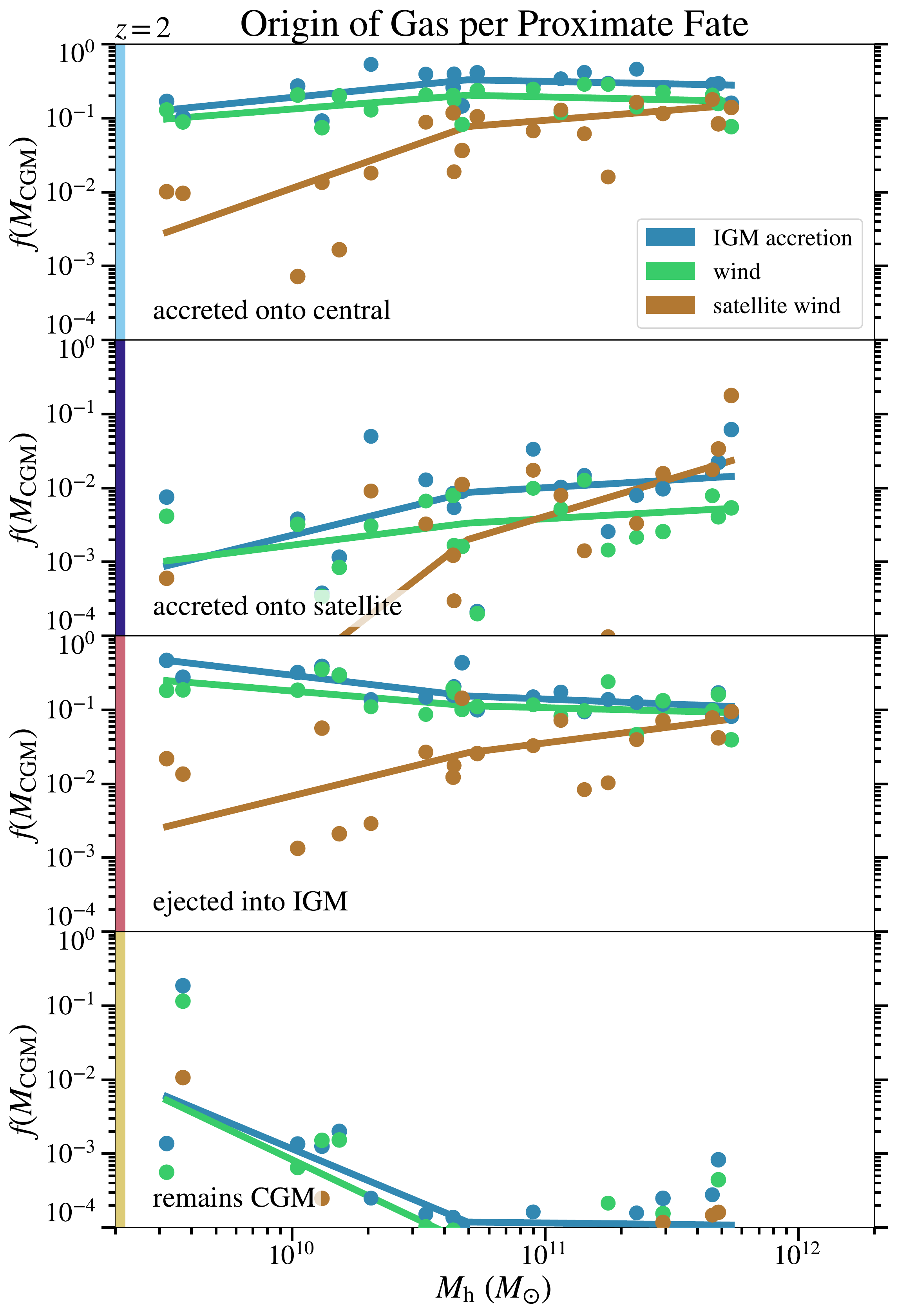}
\end{minipage} \hfill
\caption{Connection between CGM origins and proximate fates. 
Different panels correspond to different proximate fates, and different colors within each panel correspond to different origins defined in Hafen et al. (2019). 
The colored bands on the left show the colors used for proximate fates throughout the paper. 
Each point shows the CGM origin mass fraction at $z=0.25$ (left) and $z=2$ (right).  
Comparing the relative values for different colors in a single panel indicates the origin of CGM mass for a given proximate fate.
Comparing the relative values of the same color across different panels indicates the relative prevalence of proximate fates for a given origin.
For most proximate fates, the most common origin is IGM accretion, in agreement with the overall build up of the CGM (see \originp). 
The most striking exception is that, at the high-mass end and especially at $z=0.25$, the dominant origin for CGM that next accretes onto satellites is satellite wind. 
This reflects the importance of wind recycling around satellites.
}
\label{fig:CGM_origin_fate_connection}
\end{figure*}

In this section we study the relationship between the proximate fate and the origin of CGM gas.
Figure~\ref{fig:CGM_origin_fate_connection} shows CGM mass fractions at $z=0.25$ and $z=2$ for gas that has both a given proximate fate \textit{and} a given origin, as defined in \originp. 
Briefly: 
\textit{IGM accretion} is gas that has arrived in the CGM without spending $t > 30$ Myr in any galaxy;
\textit{wind} is gas that was ejected from the central galaxy;
and \textit{satellite wind} is gas that was ejected from a galaxy other than the central galaxy.
The different panels correspond to different proximate fates and the different colors correspond to different origins.
For example, the green data in the top left panel represent the fraction of total CGM mass that was ejected from the central galaxy and which will next accrete onto the central galaxy, i.e. a measure of the fraction of CGM mass from wind gas in the process of recycling.
Comparing the relative values in a single panel provides insight into the origins of a given proximate fate, and comparing the relative values of the same colors across different panels provides insight into the fates of CGM gas of a given origin.

For most proximate fates, the most common origin is IGM accretion, in agreement with the results of \originp~for the overall build up of the CGM. 
The most striking exception is for the origin of CGM gas that next accretes onto satellites, especially for $10^{11} \Msun$ and $10^{12} \Msun$ progenitor halos in the $z=0.25$ column in Figure \ref{fig:CGM_origin_fate_connection}. 
In this case, our analysis shows that the dominant origin is satellite wind. 
In other words, most of the CGM that next accretes onto satellite galaxies is recycling satellite wind.\footnote{Movies for FIRE-2 simulations available at \url{http://tapir.caltech.edu/\~sheagk/gasvids.html} illustrate this phenomenon.}

Figure~\ref{fig:CGM_origin_fate_connection} also contains an interesting result concerning the mass loading of galactic winds as they expand in the CGM. 
The third row reveals that CGM gas that will be ejected into the IGM is primarily of IGM accretion origin, not directly from galactic winds.
This indicates that winds driven by the central galaxy are significantly mass loaded in the CGM before reaching $\sim R_{\rm vir}$, echoing previous results from FIRE-1 simulations from \citet[][]{Muratov2016}. 
The effective mass loading factor in the CGM can actually be larger than suggested by this result, since much of the CGM swept up by fresh galactic winds can be ``ancient winds,'' i.e.  CGM which would appear to be of wind origin in the figure. 

\section{Discussion}
\label{sec:discussion}

\subsection{Connections to Previous Work}
\label{sec:comp_prev}

One of the most direct comparisons we can make is to \cite{Ford2014}, who also studied the origins and fates of CGM gas via a particle tracking analysis of smoothed particle hydrodynamics (SPH) cosmological simulations. 
In \originp~we discussed the extent to which the origins of CGM gas matched with those calculated by \cite{Ford2014}, finding broadly consistent results despite significant differences in the simulations.
Because \citeauthor{Ford2014} focus primarily on differentiating when gas is inside/outside a galaxy, our comparison is limited to comparing the fraction of CGM gas that is either accreted onto either the central galaxy or a satellite galaxy.
For halos with $M_{\rm h} < 10^{11.5} \Msun$ at $z=0.25$ \citeauthor{Ford2014} found that $\sim 40\%$ of the CGM gas accretes onto a galaxy, compared to $\sim30\%$ of the CGM gas that accretes at least once between between $z=0.25$ and $z=0$ at $M_{\rm h} \sim 10^{11} \Msun$ in our simulation (see proximate fates in \S \ref{fig:CGM_mass_frac_vs_Mh_CGM}). 
On the other hand for halos with $M_{\rm h} > 10^{11.5} \Msun$ at $z=0.25$ \citeauthor{Ford2014} found that only $\sim 7\%$ of the CGM gas accretes onto a galaxy, compared to up to $\sim40\%$ for some of our $M_{\rm h} \sim 10^{12} \Msun$ halos. 
While our values for lower mass halos are in broad agreement, the difference in CGM accretion for higher-mass halos is possibly a result of differences in feedback models. 
Another potentially important difference is that the simulations analyzed in \cite{Ford2014} were produced with a ``traditional'' SPH solver, and as such have suppressed cooling relative to simulations produced using hydrodynamic solvers with improved shock capturing and treatment of sub-sonic turbulence \citep[e.g][]{Keres2012, Hopkins2017}.

\cite{Tollet2019} studied the baryon cycle of galaxies and halos in the NIHAO suite of $\sim 90$ zoom-in simulations. 
On average the NIHAO halos and galaxies retain a fraction of their baryon budget similar to the FIRE-2 simulations (the baryon fractions in FIRE-2 halos are analyzed in \originp). 
\cite{Tollet2019} calculate for a given halo the fraction of the total baryonic mass accumulated over the life of the universe that is no longer inside the halo at $z=0$, and find values ranging over $\sim 0.2-0.6$.
This quantity relates to but is different from the quantities we calculate.
Our Figures~\ref{fig:CGM_gas_location} and~\ref{fig:CGM_mass_frac_vs_Mh_CGM} are closest, and contain information on the fraction of CGM mass (as opposed to total halo mass) at a given redshift (as opposed to being summed across all redshifts) that is found outside the halo at $z=0$.
For gas found in the CGM at $z=0.25$ we find that at most $\sim 40\%$ of the CGM mass is found in the IGM at $z=0$, and in most cases much less.
For gas that will be in the CGM at $z=2$ we find a significantly higher fraction of mass found outside the CGM by $z=0$ (up to $\sim 80\%$), more consistent with the values identified by \citeauthor{Tollet2019} 
\cite{Tollet2019} also briefly discuss the fraction of ejected material that originates as wind from satellite galaxies, and conclude that of the ejected material satellite winds only provide a small fraction, which we demonstrate explicitly is the case in our simulations (Figure~\ref{fig:CGM_origin_fate_connection}).

We connect our results to the galaxy-centric baryon cycle by comparing to the analysis of \cite{Angles-Alcazar2017}, who studied the origin of baryonic mass in central galaxies using a suite of six FIRE-1 simulations spanning $M_{\rm h}(z=0) \sim 10^{10} - 10^{13} \Msun$.
Among other results, \citeauthor{Angles-Alcazar2017} find that the contribution of intergalactic transfer (satellite wind that is accreted onto the central galaxy) to galactic baryons increases with increasing halo masses. 
We find a similar trend in the increase in satellite wind in the CGM that accretes onto the central galaxy with increasing halo mass in Figure~\ref{fig:CGM_origin_fate_connection}.
For $10^{12} \Msun$ progenitors the contribution of satellite wind to galactic accretion can exceed the contribution of central wind, at both $z=0.25$ and $z=2$. 
\cite{Angles-Alcazar2017} argued based on visualizations that satellites experience multiple bursts of star formation as they orbit a central galaxy, driving winds that can either accrete onto the central galaxy or recycle onto the satellite. 
Our results are consistent with this interpretation, and quantitatively demonstrate that at $z=0.25$ a comparable amount of satellite wind accretes onto the central galaxy, recycle onto satellite and remain in the CGM (as seen by comparing the orange points in the first and second rows of the left side of Figure~\ref{fig:CGM_origin_fate_connection}). 
We emphasize that accretion onto satellites is dominated by satellite wind, as opposed to gas from other origins.
However, a small fraction of the gas that accretes onto satellite galaxies is the opposite of intergalactic transfer:
wind ejected from the central galaxy that accretes onto satellite galaxies.
Observations of metal-enriched satellite galaxies may be observational evidence of this channel of the baryon cycle~\citep{Schaefer2019}.

It is useful to connect gas accretion onto satellites to star formation in satellites.
\cite{Garrison-Kimmel2019} use FIRE-2 simulations to study the star formation history of dwarf galaxies and find that satellites have a lower star formation rate than isolated galaxies, consistent with a number of observations ~\citep[e.g.][]{Baldry2006, Kimm2009, Wang2014}.
The star formation histories presented in \cite{Garrison-Kimmel2019} are broadly consistent with observations of satellite galaxies in the local group~\citep{Weisz2014, Skillman2017}.
Satellite galaxy quenching is likely related to interactions with the host halo stripping satellite ISM, a process that can be accentuated by the ejection of winds from satellites which are then more easily stripped~\citep[e.g.][]{Bustard2018}.
However, satellite galaxy quenching is not necessarily immediate or complete: \citeauthor{Garrison-Kimmel2019} find median quenching times $\gtrsim 1$ Gyr, with many $M_\star \gtrsim 10^{7} \Msun$ galaxies forming stars until $z=0$, 
consistent with observationally derived satellite quenching timescales at this mass~\citep[e.g.][]{Wetzel2015a, Fillingham2015}. 
This continued star formation may be fueled by accretion of gas from the CGM.
Of the gas in the CGM at a \textit{single} redshift (either $z=0$ or $z=2$ in our work), a gas mass greater than or equal to the total ISM mass of all satellites will accrete onto satellites (Figure~\ref{fig:CGM_mass_frac_vs_Mh_CGM} shows the fraction of CGM gas accreted onto satellites, while Figure 9 in \originp~shows the fraction of CGM gas currently in satellite ISM).
Gas that is in the CGM at other redshifts may also accrete onto satellite galaxies, making this a lower limit on the mass of gas accreted onto satellites.

\subsection{Implications for Observations}
\label{sec:comp_obs}

Our simulations that predict $\gtrsim 30\%$ of the $z=2$ CGM mass will be ejected from the halo ($R_{\rm vir}$) across all analyzed halo masses, while at $z=0.25$ the CGM mass fractions that will be ejected into the IGM by $z=0$ peak at a median $\sim10\%$ for $10^{11} \Msun$ halos (Figure~\ref{fig:CGM_mass_frac_vs_Mh_CGM}) and are typically only a few percent for $10^{12}~\Msun$ halos.
These results are consistent with observed CGM kinematics. 
At $z\sim2$ \cite{Rudie2019} measured the line-of-sight velocity for absorption systems around $10^{12} \Msun$ halos.
In 5 of the 7 galaxies with detected metal lines within the projected virial radius, \citeauthor{Rudie2019} find at least some absorption systems with centroid line-of-sight velocities that exceed the escape velocity of the halo.
\citeauthor{Rudie2019} note that these absorbers could arise from a source outside the CGM of the central galaxy, but argue that given the number of detections it is likely that at least some of the apparently unbound absorbers are associated with the CGM.
At $z <1$, absorption systems that probe the CGM in halos with $M_{\rm h} \gtrsim 10^{12} \Msun$ instead usually have line-of-sight velocities below the escape speed~\citep[e.g.][]{Tumlinson2011, Stocke2013, Borthakur2016}, consistent with much lower CGM fractions of gas ejected from our simulated $z=0.25$ $L^\star$ halos.
Moreover, \cite{Borthakur2016} find that CGM absorbers in low-redshift $\sim 10^{11} \Msun$ halos more frequently have line-of-sight velocities exceeding the escape velocity, similar to the increased fraction of ejected CGM gas for the simulated $10^{11} \Msun$ progenitors.

Identifying accretion onto galaxies has long been a goal of observational studies of the CGM~\citep[for a review see][]{Fox2017}.
We find that  $\gtrsim 80\%$ of gas with $T <10^{4.7}$ K will accrete onto a central or satellite galaxy, for halos with $M_{\rm h} \gtrsim 10^{11} \Msun$ at both $z=0.25$ and $z=2$ (Figure~\ref{fig:CGM_temp}).
Our results thus suggest that observations of low-ionization absorption systems typically probe gas that will accrete onto a galaxy, regardless of the metal content of the absorption system (which does not uniquely determine fate; Figure~\ref{fig:CGM_metallicity}). 
In $10^{12}\Msun$ halos at $z\sim0$, however, this channel is subdominant to accretion of hot gas (\S \ref{sec:temperature}).

\section{Conclusions}
\label{sec:conclusion}

We used FIRE-2 cosmological zoom-in simulations to study the fates of gas found in the $z=0.25$ and $z=2$ CGM of halos in the mass range $M_{\rm h}(z=0) \sim 10^{10} - 10^{12} \Msun$.
As with our analysis of the origins of the CGM~\citep{Hafen2019a}, we followed the full histories of individual gas elements through the duration of the simulations.
Using the particle trajectories we classified gas particles according to their proximate fate upon leaving the CGM (i.e. accretion onto the central galaxy, accretion onto a satellite galaxy, ejection into the IGM, or particles that never leave the CGM) as well as their ultimate fate (the component of the baryon cycle they reside in at $z=0$).
Our main results are as follows:

\begin{enumerate}
\item For halos with $M_{\rm h}(z=0) \sim 10^{10} - 10^{12} \Msun$, half the mass that comprises the $z=0.25$ CGM will move out of the CGM by $z=0$ (i.e. after 3 Gyr), while at $z=2$ half the mass that comprises the CGM will move out of the CGM over the course of $\lesssim 1$ Gyr (Figure~\ref{fig:CGM_gas_location}).

\item Of the CGM at $z=2$, about half end ups as central galaxy baryons (either ISM or stars) by $z=0$ in $M_{\rm h}(z=2) \sim 5\times10^{11}$ M$_{\odot}$ halos, but most of the CGM in lower-mass halos is ejected into the IGM. 
On the other hand, most of the CGM mass at $z=0.25$ remains in the CGM by $z=0$ at all halo masses analyzed. 

\item Proximate and ultimate fates are in general different because of the complex cycling of gas through galaxies and the CGM. 
For example, while most of the CGM  in $M_{\rm h}(z=2) \sim 5\times10^{10}$ M$_{\odot}$ halos is ultimately ejected into the IGM, most of it first accretes onto the central galaxy before ultimate ejection.

\item Ejection of CGM gas into the IGM depends strongly on redshift and halo mass. 
At $z=2$, when star formation is vigorous and outflows are powerful in the halos analyzed, up to $\sim 80\%$ of the CGM mass is ejected into the IGM by $z=0$, especially in lower-mass halos. 
On the other hand, the $z=0.25$  CGM mass fractions that will be ejected into the IGM by $z=0$ peak at a median $\sim10\%$ for $10^{11} \Msun$ halos and are typically only a few percent for $10^{12}~\Msun$ halos. 
These trends are qualitatively consistent with observed CGM kinematics relative to inferred halo escape velocities. 

\item Of the CGM gas that subsequently accretes onto the central galaxy in the progenitors of $M_{\rm h}(z=0)\sim10^{12}$ M$_{\odot}$ halos, most of it is cool ($T\sim10^{4}$ K) at $z=2$ but hot ($\sim T_{\rm vir}$) at $z\sim0.25$. 
This is consistent with transition from cold mode to hot mode accretion expected at this mass scale. 

\item For $M_{\rm h}(z=0) \gtrsim 10^{11} \Msun$ halos $\sim 80 \%$ of the cool/cold ($T < 10^{4.7}$ K) CGM gas will accrete onto either the central galaxy or a satellite galaxy upon leaving the CGM. 
This suggests that low-ionization absorption systems are likely to probe accreting gas. 

\item Metals have similar proximate fates as those of baryons overall, though metal-rich CGM gas is more likely to accrete onto cental galaxies at $z=0.25$ (Figure~\ref{fig:CGM_mass_frac_vs_Mh_CGM}).
Since the majority of the metals injected into the CGM by winds from central galaxies (\originp), this trend reflects the efficiency of wind recycling around low-redshift galaxies. 

\item The metallicity distributions of different fates overlap strongly, making it challenging to use observationally-inferred metallicities alone to predict the fate of absorption systems. 
Despite the substantial overlap in distributions, the median metallicity of different fates can differ by up to $\sim 1$ dex.

\item The majority of gas ejected from the CGM into the IGM is pristine gas that has never entered a galaxy (Figure~\ref{fig:CGM_origin_fate_connection}). 
This is because winds from galaxies sweep up a large amount of CGM gas as they travel to $R_{\rm vir}$ and the total CGM mass is generally dominated by IGM accretion (\originp).

\item In addition to wind recycling around central galaxies, satellite galaxies undergo frequent satellite wind recycling, even as they pass through a more massive halo (Figure~\ref{fig:CGM_origin_fate_connection}). 
Gas that accretes onto satellite galaxies can provide up to $\sim 10\%$ of the total CGM mass for $z=0.25$ $L^\star$ halos (Figure~\ref{fig:CGM_mass_frac_vs_Mh_CGM}), and is roughly half of the $T \sim 10^4$ K gas in these halos (Figure~\ref{fig:CGM_temp}).
\end{enumerate}

Overall, our particle tracking analysis demonstrates that while the CGM is affected by a complex interplay of forces and thermodynamic processes, analyzing the CGM from the perspective of fates (and origins) produces a number of important insights. 
These insights are useful both for our understanding of the complex simulation results as well as to develop a holistic framework for the interpretation of CGM observations. 
Processed data to generate the majority of the figures in this paper is available online.\footnote{\url{https://github.com/zhafen/CGM_fate_analysis}}

\section*{Acknowledgements}

We thank
Robert Feldmann,
Alex Gurvich,
Sarah Wellons,
Luke Zoltan Kelley,
Andrey Kravtsov,
and Nick Gnedin
for useful discussions.
We thank Alex Gurvich for help integrating \textsc{Firefly} into our analysis.
Zach Hafen was supported by the National Science Foundation under grant DGE-0948017. 
CAFG was supported by NSF through grants AST-1412836, AST-1517491, AST-1715216, and CAREER award AST-1652522, by NASA through grants NNX15AB22G and 17-ATP17-0067, by STScI through grants HST-GO-14681.011, HST-GO-14268.022-A, and HST-AR-14293.001-A, and by a Cottrell Scholar Award from the Research Corporation for Science Advancement.
DAA acknowledges support by a Flatiron Fellowship.
The Flatiron Institute is supported by the Simons Foundation.
JS is supported as a CIERA Fellow by the CIERA Postdoctoral Fellowship Program at Northwestern University. 
DK and TKC were supported by NSF grant AST-1715101 and by a Cottrell Scholar Award from the Research Corporation for Science Advancement.
AW received support from NASA, through ATP grant 80NSSC18K1097 and HST grants GO-14734 and AR-15057 from STScI, a Hellman Fellowship from UC Davis, and the Heising-Simons Foundation.
KE was supported by an NSF Graduate Research Fellowship.
NM acknowledges the support of the Natural Sciences and Engineering Research Council of Canada (NSERC).
This research was undertaken, in part, thanks to funding from the Canada Research Chairs program.
Numerical calculations were run on
the Quest computing cluster at Northwestern University, 
the Wheeler computing cluster at Caltech, 
XSEDE allocations TG-AST130039 and TG-AST120025, 
Blue Waters PRAC allocation NSF.1713353, 
and NASA HEC allocation SMD-16-7592.





\bibliographystyle{mnras}
\bibliography{mendeley}



\appendix

\section{Additional Pathline Visualizations}
\label{sec:more_pathlines}

\begin{figure}
\includegraphics[width=\columnwidth]{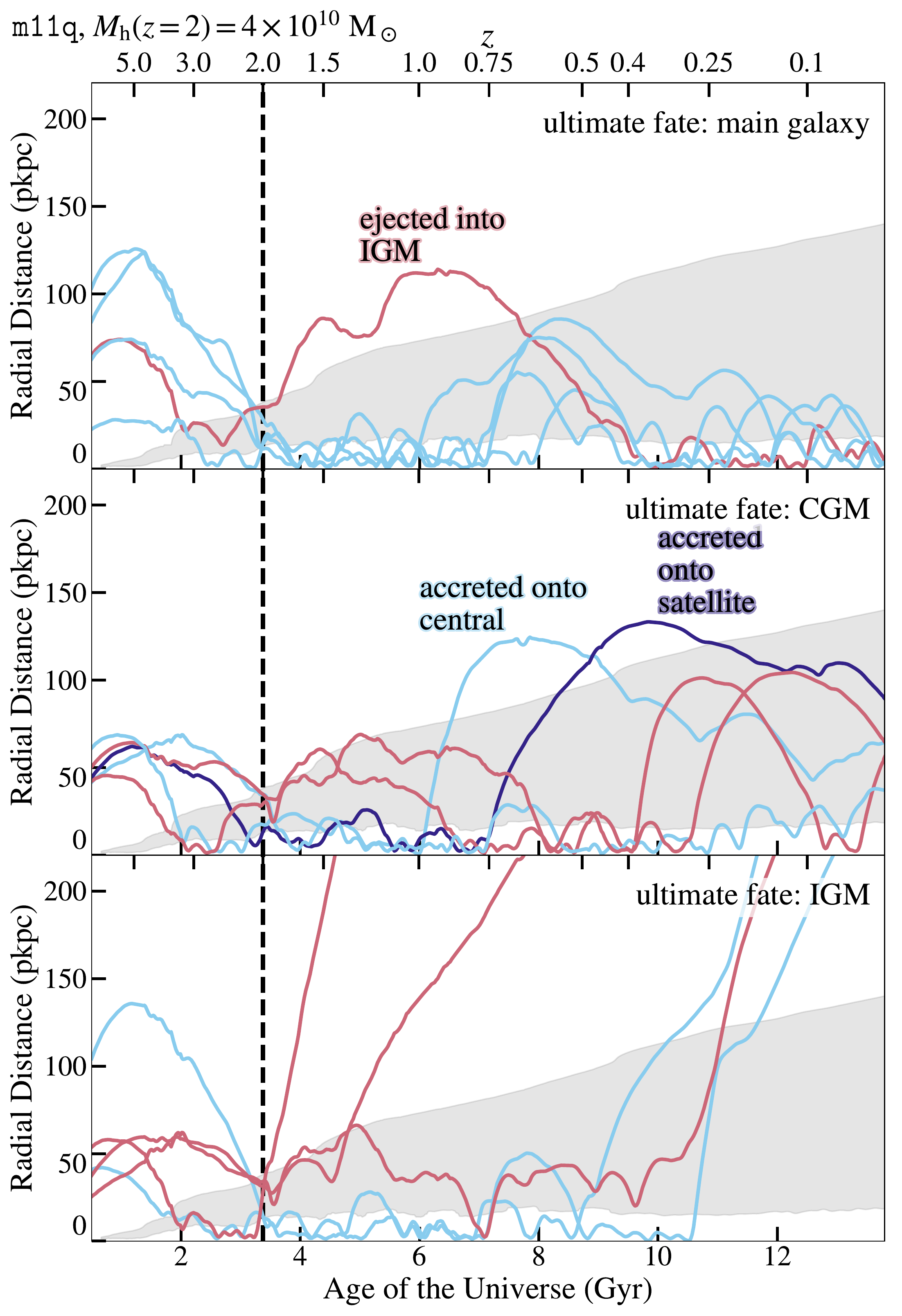}
\caption{
Same as Figure~\ref{fig:r_vs_time_eventual_m12i_CGM_snum172}, but for a representative $10^{11} \Msun$ progenitor (\texttt{m11q}).
Gas can be first lifted into the CGM, and then subsequently ejected into the IGM as part of a separate wind event.
}
\label{fig:r_vs_time_eventual_m11q_CGM_snum172}
\end{figure}

\begin{figure}
\includegraphics[width=\columnwidth]{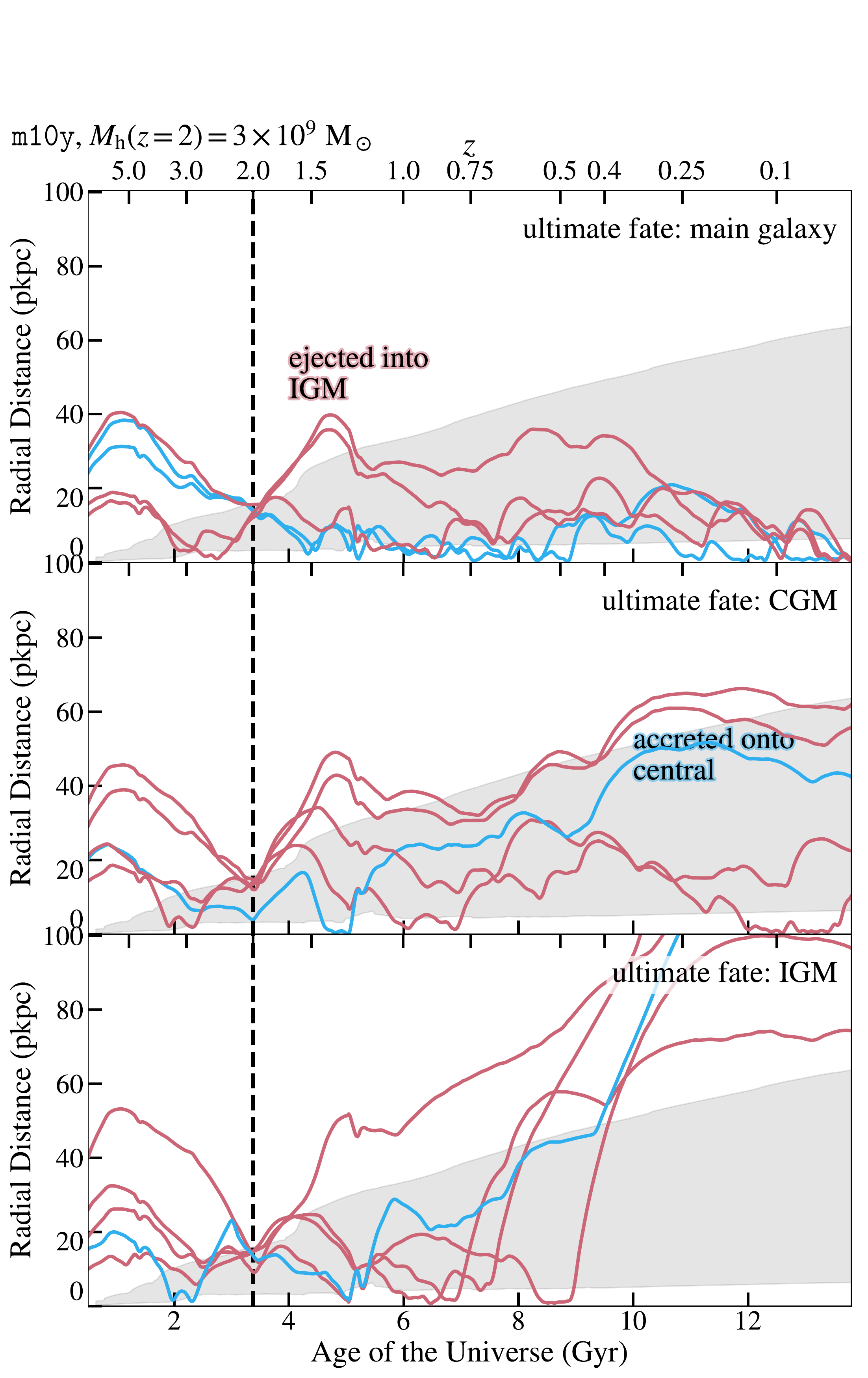}
\caption{
Same as Figure~\ref{fig:r_vs_time_eventual_m12i_CGM_snum172}, but for \texttt{m10y}.
Gas in $M_{\rm h}(z=0) \sim 10^{10} \Msun$ progenitors typically moves a smaller radial distance over a given time, relative to the halo size, and is frequently buoyed via energy injection from winds.
}
\label{fig:r_vs_time_eventual_m10y_CGM_snum172}
\end{figure}

In \S\ref{sec:particle_pathlines} we show radius vs. time for gas particles in a few example cases at $z=2$.
Here we show the same results but for a characteristic $10^{11} \Msun$ progenitor and a characteristic $10^{10} \Msun$ progenitor (\texttt{m10y}).
The CGM of $10^{11} \Msun$ is dynamic, with gas recycling far into or ejected beyond the CGM at all redshifts.
The CGM of $10^{10} \Msun$ progenitors is typically more diffuse and long-lived, with CGM gas often displaced by $\lesssim 20$ kpc over the course of $\gtrsim 5$ Gyr.


\bsp	
\label{lastpage}
\end{document}